\documentclass[sigconf]{acmart}
\usepackage{tikz}
\usepackage{amsmath}
\usepackage{microtype}
\usepackage{graphicx}
\usepackage{multicol}
\usepackage{xspace}
\usepackage{xcolor}
\usepackage{booktabs} %
\usepackage{hyperref}
\usepackage{colortbl}
\usepackage{subcaption}
\usepackage{enumitem}
\usepackage{soul}

\AtBeginDocument{%
  }

\copyrightyear{2024}
\acmYear{2024}
\setcopyright{rightsretained}
\acmConference[HPDC '24]{The 33rd International Symposium on High-Performance Parallel and Distributed Computing}{June 3--7, 2024}{Pisa, Italy}
\acmBooktitle{The 33rd International Symposium on High-Performance Parallel and Distributed Computing (HPDC '24), June 3--7, 2024, Pisa, Italy}\acmDOI{10.1145/3625549.3658685}
\acmISBN{979-8-4007-0413-0/24/06}

\newcommand{\proj}{\emph{DataStates-LLM}\xspace}

\begin{document}
\title{\proj: Lazy Asynchronous Checkpointing for Large Language Models}

\author{Avinash Maurya}
\affiliation{
    \institution{Rochester Institute of Technology}
    \city{Rochester, NY}
    \country{USA}
}
\email{am6429@cs.rit.edu}

\author{Robert Underwood}
\affiliation{
    \institution{Argonne National Laboratory}
    \city{Lemont, IL}
    \country{USA}
}
\email{runderwood@anl.gov}

\author{M. Mustafa Rafique}
\affiliation{
    \institution{Rochester Institute of Technology}
    \city{Rochester, NY}
    \country{USA}
}
\email{mrafique@cs.rit.edu}

\author{Franck Cappello}
\affiliation{
    \institution{Argonne National Laboratory}
    \city{Lemont, IL}
    \country{USA}
}
\email{cappello@anl.gov}

\author{Bogdan Nicolae}
\affiliation{
    \institution{Argonne National Laboratory}
    \city{Lemont, IL}
    \country{USA}
}
\email{bnicolae@anl.gov}
\renewcommand{\shortauthors}{Avinash Maurya et al.}

\begin{abstract}
LLMs have seen rapid adoption in all domains.
They need to be
trained on high-end high-performance computing (HPC) infrastructures and ingest massive amounts of input data. Unsurprisingly, at such a large scale, 
unexpected events (e.g., failures of components, 
instability of the software, undesirable learning patterns, etc.), 
are frequent and typically impact the training in a negative fashion. 
Thus, LLMs need to be checkpointed frequently so that they can be rolled 
back to a stable state and subsequently fine-tuned. However, given the large sizes of LLMs, a straightforward checkpointing solution that directly writes 
the model parameters and optimizer state to persistent storage (e.g., 
a parallel file system), incurs significant I/O overheads. To address this
challenge, in this paper we study how to reduce the I/O overheads 
for enabling fast and scalable checkpointing for LLMs that can be applied at high frequency (up to the granularity of individual iterations) without significant impact on the training process. Specifically, we introduce a lazy
asynchronous multi-level approach that takes advantage of the fact that
the tensors making up the model and optimizer state shards remain
immutable for extended periods of time, which makes it possible to
copy their content in the background with minimal interference during the training process. We evaluate our approach at scales of up to 180 GPUs
using different model sizes, parallelism settings, and checkpointing frequencies.
The results show up to 48$\times$ faster checkpointing and 2.2$\times$ faster end-to-end training runtime compared with the state-of-art checkpointing approaches.
\end{abstract}



\maketitle

\section{Introduction}
\label{sec:intro}

\paragraph{\bf Context.}
Large-language models~(LLMs) have seen an increasing adoption in
various domains ranging from academic and scientific research to
industrial applications. They have been traditionally used for creative 
text generation, prompt completion, comprehension, and summarization, etc.
Additionally, recent initiatives such as LLMs for science (e.g., DeepSpeed4Science
\cite{songDeepSpeed4ScienceInitiativeEnabling2023}) are beginning to
explore use cases that involve specialized domain-specific languages
for tasks such as genome sequencing, protein structure prediction,
equilibrium distribution prediction, etc. The versatility and
democratization of LLMs have led to an unprecedented scale of development
and discovery across multiple fields.

In a quest to improve the quality of large language models (LLMs), the
size of the training data and the size of the LLMs are rapidly
increasing. LLMs are routinely made of billions of parameters and
there are predictions that they will reach the trillion scale in the near
future, e.g., Google Switch-C (1.6T)~\cite{google-switch}, WuDao 2.0 (1.75T)~\cite{zeng2022glm}, and M6-10T~\cite{lin2021m6}. 
Under such circumstances, LLMs need to be trained in a
scalable fashion on high-performance computing (HPC) machines comprising a large number of compute nodes and GPUs. Despite advances in technologies that enable LLM
training to scale (hybrid data-, pipeline- and tensor parallelism,
sharding of model parameters and optimizer state, layout and
communication optimizations, etc.), training remains a resource and
time-intensive task: LLMs often require weeks if not months to either
be trained from scratch (also referred to as pre-training) or be
fine-tuned for specialized tasks.

\paragraph{\bf Motivation: Checkpointing as a Fundamental Primitive.}
During such a long runtime involving a large number of components,
unexpected events are frequent and can
have devastating consequences. 
For example, due to the tightly coupled nature of distributed training of LLMs, hardware failures, software bugs, or communication timeouts, can occur, which may lead to globally corrupted states even if they affect a small number of components. 
Unicron~\cite{he2023unicron}, a recent effort from 
Alibaba, highlights a 43.4\% failure rate of resource-intensive LLM training, 
out of which 37\% were hardware failures, while the remainder 73\% could be fixed by system restarts. In both cases, a checkpoint is needed to effectively resume the LLM training.

Even in the absence of failures, the training
can take an undesirable trajectory that leads to dead-ends, e.g., slow
or no convergence, undesirable learning patterns that need to be
``unlearned'', instability, etc. For example, loss spikes are one type of an undesirable trajectory. They were reported by PaLM~\cite{chowdhery2023palm} and GLM-130B~\cite{zeng2022glm} and were observed during the training of popular models such as BLOOM-175B and OPT-175B.
Since they are hard to predict and defend
against, the only viable strategy is to roll back to a past checkpoint 
and try an alternative strategy, such as skipping over problematic
mini-batches or reorganizing the model, e.g., by switching some parameters to 
higher precision or different floating point representation.

Additionally, checkpointing of
intermediate states during the training is a fundamental pattern used
in several other scenarios: understanding the evolution of the
learning patterns captured by the model, continuous testing of
alternatives without disturbing production deployments, switching
between divergent model states based on Reinforcement Learning from
Human Feedback (RLHF).

\paragraph{\bf Challenges and Limitations of State of the Art.}
Widely used deep-learning models (ResNet~\cite{resnet}, VGG~\cite{vgg-paper},
etc.) of moderate
sizes, i.e., hundreds of MBs, and their associated optimizer state
typically fit in the memory of a single GPU. In this case, data
parallelism is often enough to scale the training, which means that
it is enough to checkpoint a single model replica by gathering the
relevant state from a single GPU. On the other hand, LLMs are sharded
across a large number of GPUs, which means that a checkpoint needs to
gather distributed data structures. Such an operation involves
much larger sizes, i.e., in the order of hundreds of GBs. 
Therefore, synchronous checkpointing solutions, e.g. default checkpointing implemented in DeepSpeed~\cite{rasleyDeepSpeedSystemOptimizations2020}, that block the training until the model state is captured to stable storage incur high runtime overheads. 
Alternatively, one may use a multi-level asynchronous checkpointing solution
that copies the model state to a fast memory tier and then flushes
it from there to slower tiers in the background without blocking the
training. In general, this is a widely used solution in the HPC
community that successfully reduces the runtime overheads compared
with synchronous checkpointing. However, it is not straightforward to
implement this approach in the context of LLM training for two reasons.
First, there is simply not enough free memory on the GPUs to
hold a full copy of the checkpoint shards, due to which it is not possible to
benefit from the high GPU memory bandwidth to reduce the overhead
of creating a blocking copy. Second, while it is possible to create
the copy directly on the host memory (e.g. TorchSnapshot~\cite{TorchSnapshot}, TorchLightning~\cite{WelcomePyTorchLightning}, CheckFreq~\cite{mohanCheckFreqFrequentFineGrained}),
this involves data transfers that are an order of magnitude slower
and subject to competition due to shared PCIe links between multiple
GPUs and the host memory.  Ultimately, this results in significant
overheads that reduce the benefit of asynchronous checkpointing
to the point where it is not significantly faster as compared to
synchronous checkpointing approach. To put this in perspective, despite 
the availability of high speed links 
(50+~GB/s network and 25+~GB/s PCIe),
the LLM checkpointing throughput is far from saturating the 
link capacity (e.g., REFT~\cite{wang2023reliableREFT} reports 
38\% saturation), and often drops as low as a few GB/s (e.g.,
Nebula~\cite{ziqiwangOptimizeCheckpointPerformance2023}, Microsoft's 
DeepSpeed closed-source implementation of asynchronous checkpointing
reports 1-4 GB/s).

\paragraph{\bf Key Insights and Contributions.}
In this paper, we propose \proj, a novel asynchronous checkpointing technique
that overcomes the limitations of the aforementioned state-of-the-art approaches. Our key idea is to leverage the observation that the model parameters
and optimizer state remain immutable for extended periods of time during
an iteration (i.e., during the forward pass and backward pass) and are updated 
in bulk at specific points. Specifically, we can copy the model state 
(parameters, optimizer state) during the forward and backward pass
from the GPU to the host memory without blocking the training
iteration. At the same time, we can hide the overhead of contention 
for the memory and storage tiers and guarantee the consistency of
checkpoints asynchronously once the checkpointing data is available
on the host memory. We summarize our contributions as follows:

\begin{enumerate}

\item We perform a gap analysis that highlights the checkpoint
sizes, load-balancing among the checkpoint shards corresponding
to 3D parallelism, and when the LLM model parameters and optimizer
state remain immutable during each training iteration. This
analysis is essential in shaping our contribution (\S~\ref{sec:motivation}).

\item We introduce a series of key design principles, i.e., hybrid flushing of 
GPU model/optimizer shards to host memory, lazy copy that overlaps
with the intervals during which the LLM remains immutable,
streamlined multi-level flushing to persistent storage, and
asynchronous consolidation of model/optimizer shards (\S~\ref{sec:design}). 

\item We discuss an architecture
that integrates these design principles into widely used LLM training
runtimes, namely DeepSpeed and Megatron (\S~\ref{sec:arch}).

\item We design and implement the components of the architecture,
insisting on details related to high-performance aspects, such as,
efficient data movements and serialization of LLM shards, orchestration
of background parallelism, bridging between high-level abstractions 
in Python and low-level C++ implementation, coordination and consistency
(\S~\ref{sec:impl}). 

\item We evaluate our implementation in 
a series of extensive experiments in which we train large LLMs (up to
70B parameters) on modern HPC systems (512 nodes, each consisting of four A100 40GB GPUs).
We show significant speed-up of end-to-end runtime and up to 4$\times$ higher
checkpointing throughput in a variety of configurations (\S~\ref{sec:evaluation}).

\end{enumerate}

\begin{figure*}[t]
    \centering
    \includegraphics[width=\linewidth]{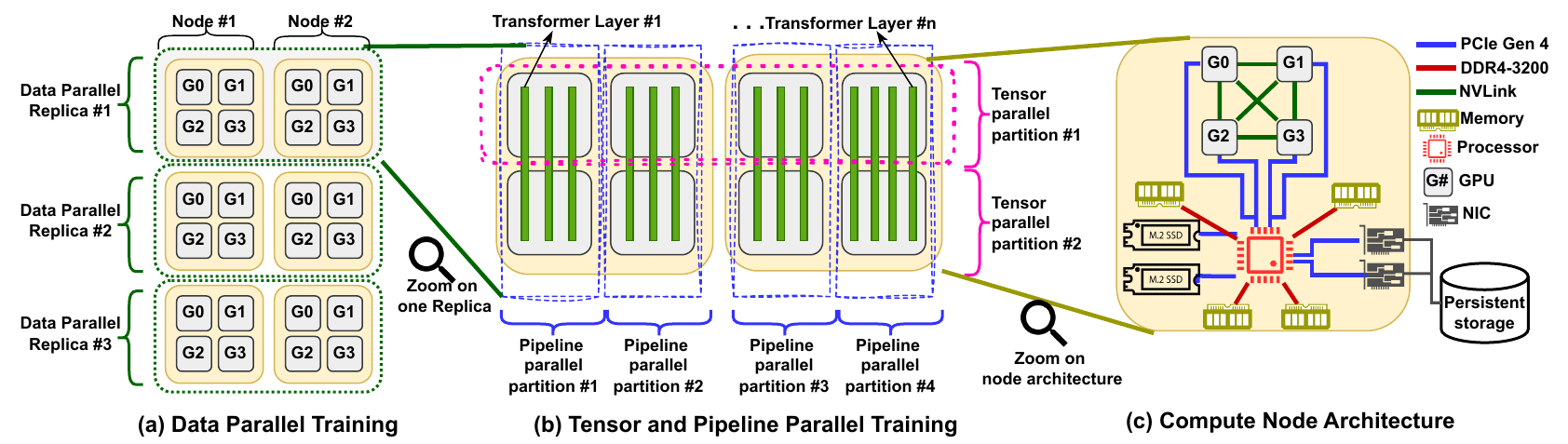}
    \caption{Data, pipeline, and tensor parallel runtime training. Compute node configuration consisting of four A100-40GB GPUs.}
    \label{fig:arch}
    \Description{
    A diagram that illustrates data, pipeline and tensor parallelism
    mapped on a typical compute node architecture consisting of four A100-40GB GPUs.}
\end{figure*}

\paragraph{\bf Limitations of the Proposed Approach.}
By leveraging the fact that the LLM remains immutable during a significant
part of each training iteration, we can perform lazy device-to-host copies 
of the tensors that make up the LLM model state, which reduces the time
each iteration is 
blocked while waiting for device-to-host I/O related to checkpointing to finish. 
This accelerates the training iterations 
during which a checkpoint is taken, but at the cost of accumulating
checkpointing data on the host memory faster, especially for high
checkpoint frequencies. Thus, if the asynchronous flushes of the 
checkpointing data from the host memory to the lower-level storage 
tiers, e.g., node-local NVMe storage and parallel file systems (PFS), are not fast 
enough to keep up with the device-to-host lazy copies, this will
eventually become a bottleneck. In this case, our approach needs
to be complemented with other techniques, e.g., compression, 
for reducing
the bottleneck caused by the flushes. Nonetheless, even 
under such circumstances, our approach will still exhibit less
overhead than other state-of-the-art LLM checkpointing approaches, 
albeit the difference will be smaller.

\section{Background}
\label{sec:background}

\subsection{Data Parallelism} 

Data parallelism is the most widely used technique to accelerate the training 
of deep learning models~\cite{DBLP:journals/pvldb/LiZVSNLPSVDC20}. It creates replicas of 
the learning model on multiple workers, each of which is placed on a different device and/or
compute node, as illustrated in Figure~\ref{fig:arch}(a). The input data is randomly shuffled and partitioned among the workers at each epoch. 
During the forward pass, the workers simply process their
mini-batches from the partition of their dataset in an embarrassingly 
parallel fashion. Then, during the backward pass, the model parameters 
are updated based on the average gradients of all replicas (instead of the
local gradients), which effectively synchronizes all replicas to learn
the same patterns from all partitions. 
Data parallelism leads to accelerated training because the partitioning of
the input data results in fewer iterations per epoch. 

\subsection{Pipeline and Tensor Parallelism}
Pipeline and tensor parallelism
are two complementary techniques that enable the training of large learning models 
that do not fit in the memory of a single GPU. Pipeline parallelism leverages the 
idea that learning models can be split into stages, each of which can placed on
a separate GPU. Then, the forward and backward pass corresponding to different 
mini-batches can be overlapped by activating all stages in parallel: as soon as 
the forward pass of one mini-batch has been moved to the next stage, another mini-batch
can be processed in the current stage. This idea applies similarly to the backward
pass but in reverse order: as soon as the backward pass of one mini-batch has been moved
to the previous stage, another mini-batch can be processed in the current stage~\cite{huang2019gpipe}.
Tensor parallelism leverages the idea that even individual layers and tensors can
be sharded and distributed horizontally across multiple GPUs. Figure~\ref{fig:arch}(b)
illustrates these ideas for an example decomposition of an LLM consisting of $n$ layers into 
multiple pipeline parallel (highlighted by the vertical dotted blue box) and tensor parallel 
(denoted by the horizontal dotted magenta box) shards. Nvidia Megatron-LM is a prominent example
of the LLM framework that is widely adopted in practice and offers configurable mechanisms to partition 
the model in pipeline and tensor parallel modes. The trade-off in this case is that 
the computations on the stages and shards are tightly coupled and distributed at fine 
granularity among the GPUs, which introduces the need for frequent communication that 
is subject to overheads. Amongst data, pipeline, and tensor parallel approaches model training, 
the tensor parallel approach is the most communication-intensive approach since it requires intra-layer 
interaction. Therefore, if tensor-parallelism cannot be completely avoided for a given model configuration, 
it is typically configured to use node-local GPU resources, thereby exploiting fast node-local fabrics such 
as NVLinks~\cite{a100-nvlink}.
On a typical A100 GPU compute node, illustrated in Figure~\ref{fig:arch}(c), the degree of 
tensor parallelism should not exceed the number of node-local GPUs in order to take advantage of fast 
600~GB/s NVLinks to mitigate communication overheads. The combination of data parallelism, pipeline parallelism, and tensor parallelism is often called \emph{3D parallelism}.

\begin{figure*}[t]
    \centering
    \includegraphics[width=0.99\linewidth]{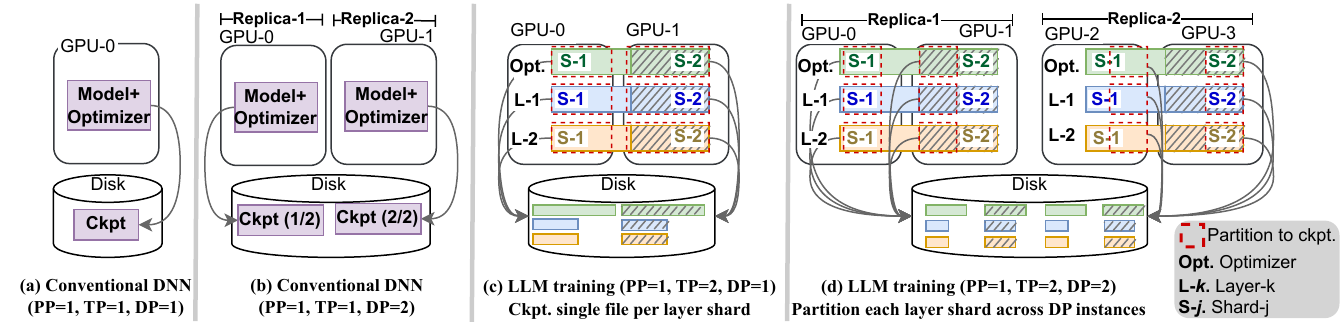}
    \caption{Sharding of checkpoints during training of conventional DNNs and LLMs for different degrees of pipeline (PP), tensor (TP), and data (DP) parallelism.}
    \label{fig:ckpt-sharding}
    \Description{
    A diagram that illustrates the composition of LLM checkpoints as they 
    are captured from distributed shards corresponding to data, pipeline and tensor parallelism. To enable efficient I/O, different layer shards and optimizer shards are written to different files.
    }
\end{figure*}

\subsection{State Sharding to Eliminate Redundancy of Data-Parallel Replicas}
Data parallelism introduces high redundancy
in maintaining independent model replicas. This can be exploited to maintain a single
replica across all workers, 
where each replica
is responsible for the management of a distinct 
shard. Then, when a worker needs to access a full model, it needs to obtain all missing
shards from the rest of the workers. Just like in the case of model parallelism, such an approach
sacrifices performance, due to extra communication overheads, for improving memory efficiency.
A prominent example that implements this idea is DeepSpeed~\cite{rajbhandariZeROMemoryOptimizations2020}
which is widely used for training
LLMs in combination with Megatron~\cite{shoeybiMegatronLMTrainingMultiBillion2020}. 
DeepSpeed offers a set of incremental optimization stages: 
stage-1, stage-2, and stage-3, which correspond to sharding the optimizer state, gradients, and 
model parameters 
across all data parallel ranks, respectively. 
DeepSpeed also offers additional tunable optimizations
such as out-of-core management of shards using the host memory for swapping.

\subsection{Implications of State Sharding on Checkpointing}
For conventional DNN models, the state captured in the checkpoint (typically
model parameters and optimizer state) is usually serialized as a single file,
as depicted in Figure~\ref{fig:ckpt-sharding}(a). When using data parallelism,
since there are many identical DNN model replicas available, it is possible
to split the model into shards and parallelize the I/O by ensuring each
worker captures and flushes a different shard as a separate file, as shown
in Figure~\ref{fig:ckpt-sharding}(b). This approach is adopted by
DeepFreeze~\cite{nicolaeDeepFreezeScalableAsynchronous2020}, TorchSnapshot~\cite{TorchSnapshot}, and LightCheck~\cite{lightCheck}.
In the case of LLMs, sharding can be exploited even without data parallelism
to enable parallel writes of different layers into different files, as shown 
in Figure~\ref{fig:ckpt-sharding}(c). Finally, this can be complemented
by another level of sharding when data parallelism is added, as shown
in Figure~\ref{fig:ckpt-sharding}(d). By default, the DeepSpeed runtime
implements the latter case, which results in a large number of shards
being stored in separate files. On many HPC systems, this provides the best
I/O performance especially for parallel file systems. However, 
it also raises the problem of managing a large number of shards and
potential metadata bottlenecks~\cite{IOAgg-ISPDC23}. In this work, we assume that the default
DeepSpeed strategy is to serialize the LLM checkpoint shards into separate 
files while leaving the question of how to find better file aggregation layouts as
future work.

\subsection{Problem Formulation}
For the scope of this paper, we only focus on scenarios considering 3D parallelism combined with stage-1 (optimizer partitioning across data-parallel ranks), which corresponds to a configuration in which DeepSpeed and Megatron were successfully used to train the largest LLM models, such as BLOOM~\cite{workshopBLOOM176BParameterOpenAccess2023} (up to 175 billion parameters). 
Our goal is to design scalable multi-level asynchronous checkpointing solutions that: (1) capture a globally consistent 
checkpoint of LLMs that includes all shards of all GPUs corresponding to both the model parameters and the 
optimizer state (which is needed to successfully restart the training); (2) maximize the checkpointing throughput 
in order to reduce the amount of time during which the training is blocked by checkpointing; and (3) minimize the contention 
for resources and interference between the training and the overlapped background data transfer tasks 
for reducing
the end-to-end training duration.

\section{Related Work}
\label{sec:related}

\subsection{Checkpointing in Deep Learning}
Checkpointing techniques have been extensively explored in the specific context of deep learning for minimizing the I/O overheads on training. Systems such as CheckFreq~\cite{mohanCheckFreqFrequentFineGrained} aim at performing fine-grained iteration-level checkpoints and overlap checkpoint flushes with the training phases, but do not support checkpointing in pipeline parallel training setups and are
inefficient in utilizing the available network and PCIe interconnect and memory subsystems, showing only up to 40\% peak efficient 
checkpointing throughput
across data-parallel replicas. Approaches such as DeepFreeze~\cite{nicolaeDeepFreezeScalableAsynchronous2020}, TorchSnapshot~\cite{TorchSnapshot}, and LightCheck~\cite{lightCheck} attempt to mitigate the checkpointing overheads by both overlapping transfers with training and partitioning checkpoints across data-parallel replicas, but do not support hybrid pipeline, tensor, data-parallel training setups. 

\subsection{Checkpointing for LLMs}
Several recent efforts specifically target checkpointing for LLMs and focus on efficient asynchronous 2-phase CPU-based snapshotting and lazy persistence. 
However, the reported checkpointing throughputs are far from saturating the network (50+~GB/s
and PCIe (25+~GB/s) links. For example, Gemini~\cite{wang2023gemini} reports 3.13~GB/s checkpointing throughput (9.4~GB shard of GPT-100B takes about 3 seconds for checkpointing).
REFT~\cite{wang2023reliableREFT} reports 38\% PCIe bandwidth utilization at 6~GB/s,
while TRANSOM's checkpointing engine~(TCE)~\cite{wu2023transom} reports achieving a throughput of $\sim$1.2~GB/s. Nebula~\cite{ziqiwangOptimizeCheckpointPerformance2023}, which is Microsoft's DeepSpeed closed-source implementation of asynchronous checkpointing and is
only available on the Azure cloud, reports achieving 1-4~GB/s (GPT2-XL checkpoint of 20.6~GB takes 5 seconds to checkpoint). These results hint at significant gaps in existing checkpointing techniques for LLMs.

\subsection{High-Performance Checkpointing Runtimes}
HPC workloads have widely adopted checkpointing runtimes for resilience. User-transparent runtimes, e.g., BLCR~\cite{hargrove2006berkeley} and DMTCP~\cite{ansel2009dmtcp}, capture the entire state of all processes distributed across multiple nodes, which is exclusively used for restarting from failures. GPU-based transparent checkpointing runtimes such as CheCUDA~\cite{takizawa2009checuda} and NVCR~\cite{nukada2011nvcr} provide similar functionality for capturing GPU-based working state of the application. While these approaches are transparent, they incur higher checkpointing overhead because the entire state of the application (including non-critical data structures) is captured and flushed to disk. Application-level checkpoint-restart runtimes such as VELOC~\cite{nicolaeVeloCHighPerformance2019, VELOCGPU-HiPC22, LineageComp-HIPC23, GPUPrefetch-HPDC23} and FTI~\cite{bautista2011fti, parasyris2020checkpoint-fti-gpu} require the application to mark critical data structures necessary to restart application from failures for both CPU-only and hybrid CPU-GPU applications. Canary~\cite{canary-sc22} supports containerized checkpointing. However, none of these runtimes exploit the immutable phases of LLM training to optimize checkpointing by overlapping the checkpointing phase with the training phase.

\subsection{I/O Optimizations in Data Movement and Checkpoint Runtimes}
Data-movement and checkpoint engines in HPC such as ADIOS~\cite{godoy2020adios}, VELOC~\cite{nicolaeVeloCHighPerformance2019, VELOC-MASCOTS21}, and FTI~\cite{bautista2011fti} support efficient asynchronous data movement through multi-level cache hierarchy. VELOC~\cite{VELOCGPU-HiPC22}, for instance, reserves a pinned cache on both the device (GPU) and host memory for buffering checkpoints in an overlapping fashion with the application execution. However, given the large device memory required for LLM training, the GPU does not have enough spare capacity to even hold a few tensors that need to be checkpointed; thereby compelling runtimes to use host memory as the fastest memory tier to cache/buffer checkpoints from. Furthermore, unlike conventional DNNs where the size of the input dataset is typically larger than the model states (and therefore checkpoints), in the case of LLMs, the model is usually larger than the micro-batches consisting of a few thousand integer-based tokens. Therefore, as highlighted in Gemini~\cite{wang2023gemini}, the available pool of host memory is generally large enough to accommodate both the next subset of prefetched input micro-batches and LLM checkpoints.

\begin{figure}[t]
    \centering
    \includegraphics[width=\linewidth]{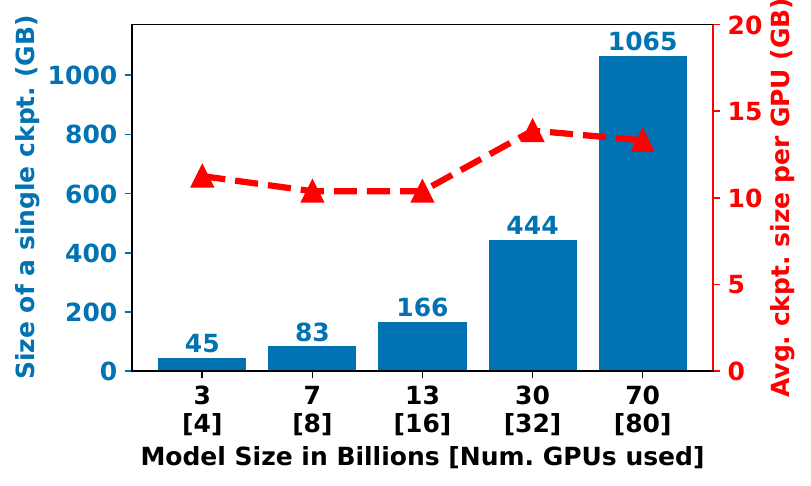}
    \caption{Aggregate checkpoint sizes of different model sizes and average checkpoint size per GPU.}
    \label{fig:agg-ckpt-size}
    \Description{
    A diagram that motivates the need for scalable checkpointing techniques:
    checkpoint sizes are quickly with increasing number of parameters and 
    reach the order of TBs, resulting in significant I/O overheads.
    }
\end{figure}

\section{Analysis of LLM Checkpointing Behavior}
\label{sec:motivation}

\subsection{LLM Checkpoint Sizes and Load Balancing}
Unlike the case of lightweight optimizers such as stochastic-gradient descent~(SGD)~\cite{ruder2016overview},
which are widely used in conventional DNN models, LLMs adopt advanced 
adaptive learning rate optimizers such as Adam~(Adaptive momentum estimation)~\cite{kingma2014adam}.
Such optimizers need to store additional state information (momentum, variance, gradients), which leads to an explosion of the optimizer state size. Unfortunately, this state information cannot be simply left out of the checkpoint as it is essential for a successful restart of the training process. Coupled with the already large number of LLM parameters (billions),  the overall checkpoint size becomes massive. Even worse, while the size of checkpoints grows proportionally to the number of transformer layers, it grows quadratically with respect to the number of hidden dimensions~\cite{rajbhandari2021zero}. To study
this effect, we ran a series of experiments (the setup is explained in detail in
\S~\ref{sec:setup}) that use DeepSpeed to train the models listed in Table~\ref{tab:models}. The results are depicted in Figure~\ref{fig:agg-ckpt-size}.
As expected, the checkpoint sizes quickly grow to large sizes and exhibit
similar checkpoint size per GPU for different model sizes, hinting at the fact
that DeepSpeed achieves good load-balancing among the shards as highlighted by 
the minor y-axis.

\begin{figure}[t]
    \centering
    \includegraphics[width=\linewidth]{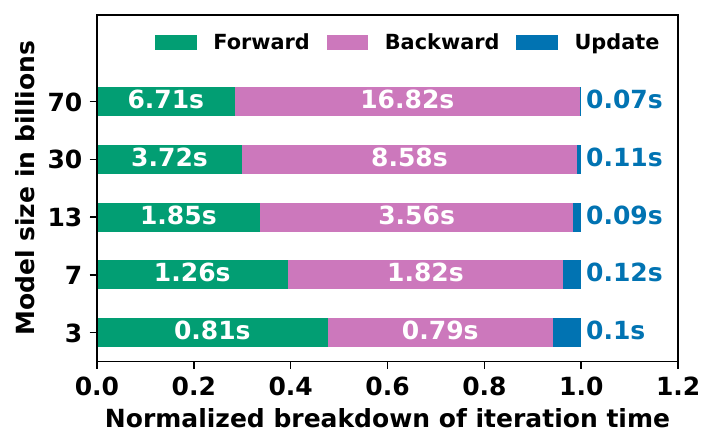}
    \caption{Different iteration phases. Model and optimizer states are immutable during 
    forward and backward passes.}
    \label{fig:diff-phases-time}
    \Description{
    A diagram that shows the breakdown of a training iteration in terms
    of durations of forward pass, backward pass and update phase. With
    increasing number of parameters, the forward and backward pass
    take increasingly longer time to finish vs. the update phase.
    }
\end{figure}

\begin{figure*}[t]
    \centering
    \includegraphics[width=0.95\linewidth]{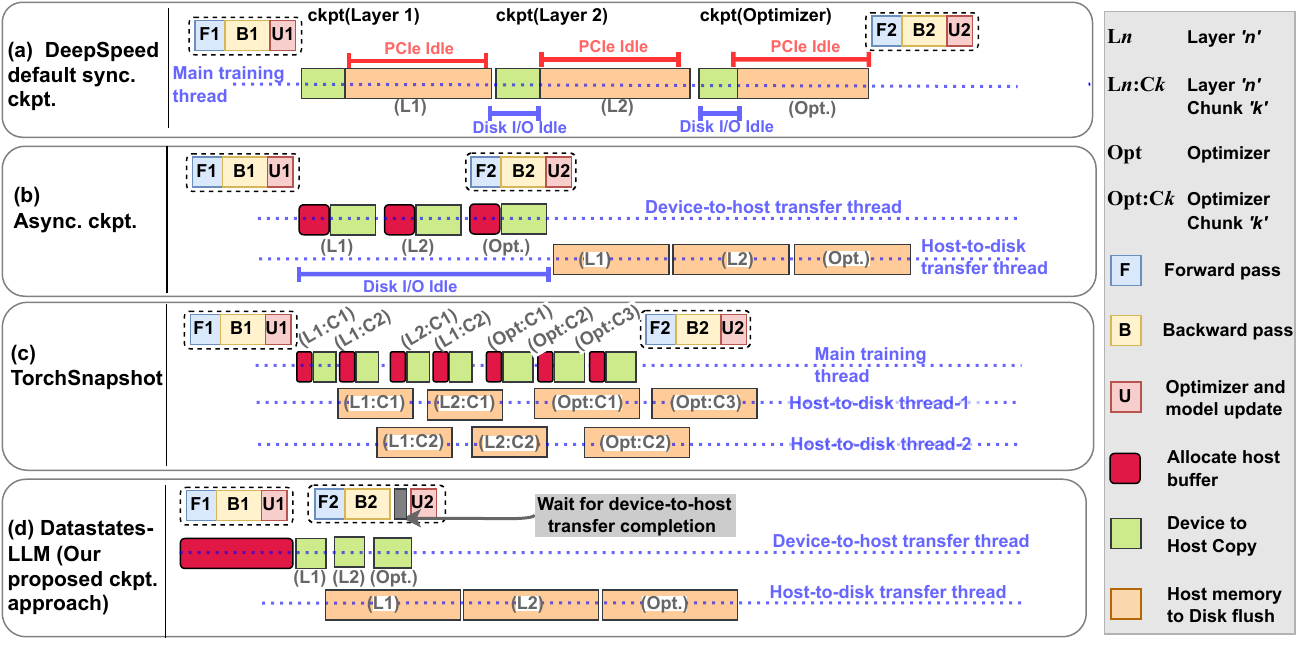}
    \caption{Overlapping LLM training with checkpointing using different approaches.}
    \label{fig:overlap-llm-flow}
    \Description{
    A diagram comparing the behavior synchronous checkpointing,
    asynchronous checkpointing that performs blocking copies 
    from the GPU to the host memory, then flushes the 
    checkpointing data asynchronously to stable storage,
    TorchSnapshot, which improves on the previous approach
    by using multiple asynchronous I/O threads, and our approach,
    which takes advantage of immutability during forward pass
    and backward pass to perform asynchronous copies from
    the GPU to the host, effectively reducing device-to-host
    I/O overheads.
    }
\end{figure*}

\subsection{Immutability of Model Parameters and Optimizer States During Each Iteration}

We also 
study the behavior of each training iteration at fine granularity by breaking
down the runtime into forward pass, backward pass, and update duration. The
results are shown in Figure~\ref{fig:diff-phases-time}. 
We observe that
regardless of the model size, the forward and backward passes take up the majority of
the training iteration duration. 
In addition to the increasing computational complexity involved in training larger models, 
the long iteration duration can be attributed to
operations, such as, send/recv of activations and gradients (pipeline and
tensor parallelism) and gradient all-reduce (data parallelism), are expensive
and become a bottleneck. With increasing the LLM model size, they get amplified
and lead to a negligible update phase. Fortunately, this situation presents
an opportunity that can be leveraged to our advantage. First, both the 
model parameters and the optimizer state remain immutable during both
the 
forward and backward passes. 
Thus, any copies from the
GPU memory to the host memory can be issued asynchronously during the 
forward pass and the backward pass without causing coherency issues. Second, such copies utilize the PCIe link between the GPU and the host, which is different
from the communication links (i.e., NVLink~\cite{a100-nvlink} and GPUDirect RDMA~\cite{potluri2013efficient}) between GPUs and between the compute nodes 
that are used for communication during training. Thus, asynchronous copies do 
not compete for bandwidth with the forward and the backward passes and therefore 
they do not cause interference. 

\section{DataStates-LLM: System Design}
\label{sec:designandimpl}

\subsection{Design Principles}\label{sec:design}
Based on the observations outlined in \S~\ref{sec:motivation}, we 
introduce a series of high-level design principles that we adopt in \proj 
to mitigate the limitations of
state-of-art LLM checkpointing runtimes.

\paragraph{\bf Coalescing of GPU Model/Optimizer Shards to Host Memory:}
Conventional asynchronous multi-level checkpointing techniques (as implemented in the related works mentioned
in \S~\ref{sec:related}) move the checkpoints one-at-a-time through the storage levels: first they allocate
host memory to hold the checkpoint, then they capture the checkpoint on the host memory by performing a GPU-to-host 
copy, then they asynchronously flush the checkpoint from the host memory to persistent storage. If another checkpoint
request arrives before the previous checkpoint is finished flushing, it will be blocked waiting for the flushes to
complete. For small learning models that fit in the memory of a single GPU, such an approach works reasonably well
because all model parameters and the optimizer state can be captured at once in a single file. However, the combination 
of 3D parallelism and optimizer state sharding targeted by our checkpointing scenario results in many independent shards 
per GPU that correspond to both the model parameters and the optimizer state. Eventually, each of these shards needs to be flushed to persistent storage, typically as a separate file, as illustrated in Figure~\ref{fig:ckpt-sharding}(c).

In this case, conventional asynchronous multi-level
approaches would serialize the checkpointing of the shards.
For example, if we consider three shards in a checkpoint, two of which correspond to layers $L1$ and $L2$ and the third corresponds to the optimizer state
shard, then only the flushing of the optimizer state shard will overlap with the next iteration (forward pass,
backward pass, and updates), while the rest of the operations (allocate, copy, flush $L1$; allocate, copy, flush $L2$;
allocate, copy optimizer state) are serialized. This severely degrades the performance of asynchronous checkpointing
to the point where it may become slower than synchronous checkpointing. 
To optimize and extend the conventional asynchronous multi-level 
checkpointing approach for multi-layered LLMs, the following approach, illustrated in Figure~\ref{fig:overlap-llm-flow}(b), can be used --- all the three shards in the checkpoint ($L1$, $L2$, and optimizer) can be first \emph{snapshotted} quickly using device-to-host copies, which will block the training for all layers except the snapshot of last layer, which can be overlapped with the next training iteration. Once the snapshot of all layers involved in the checkpoint is complete, they can be persisted through asynchronous flushes from host to disk. However, even such an advanced asynchronous approach slows down training due to slow host memory allocation and transfers (as evaluated in Figure~\ref{fig:vary-ckpt-iter-e2e-13B}).
To mitigate this issue, we propose
three optimizations. First, we pre-allocate enough host memory to hold all shards on the host memory. This pre-allocated
memory will be reused for all checkpoint requests, effectively eliminating the allocation overheads for all shards,
both belonging to the 
same and different checkpoints. 
Second, we pre-pin the allocated host memory,
which accelerates GPU-to-host data transfers, again for all shards of both the same and different checkpoints. Third,
we coalesce the copies of the shards to host memory, which eliminates the need to wait for the flushes of
the shards belonging to the same checkpoint to finish before initiating more GPU-to-host copies.

\paragraph{\bf Lazy Non-Blocking Copies Overlapping with Forward and Backward Pass:}
We leverage a key observation that the model and optimizer shards on each GPU remain immutable during the forward pass and the backward
pass, and are updated later in bulk (typically through \verb|optimizer.step()| for optimizers such as Adam). 
Therefore, unlike conventional asynchronous multi-level checkpointing techniques,
there is no need to block the next training iteration until a full copy of the checkpoint is available on the host memory.
Instead, we allow the next training iteration to start immediately after the checkpoint request, and proceed to
copy the shards to the host memory while the forward pass and the backward pass are progressing in parallel. Only
when the update phase is about to begin, if the shard copies on host memory are not finished, then we delay the
update phase until they are finished. Furthermore, the flushes from the host memory to persistent storage are
also allowed to overlap with the update phase. It is for this reason that we refer to our technique as ``lazy'' 
non-blocking copies: in effect, we reduce the duration of blocking the training by postponing the wait for as
much as possible until there is a risk for consistency issues. An example is illustrated in 
Figure~\ref{fig:overlap-llm-flow}(d): the forward and backward pass of the second iteration $F2$ and $B2$ proceed
immediately after the first iteration has finished, at which point a checkpoint request was issued. They overlap
with the GPU-to-host copies. The update phase $U2$ is delayed until the GPU-to-host copies have finished, thereby blocking the application. Meanwhile, the previously captured checkpoints on the host are asynchronously flushed to persistent storage. Finally, if the host memory that is reserved for checkpointing is full, then the next checkpoint request
needs to wait for previous tensors to get evicted from the host memory after they are flushed
to the persistent storage, e.g., node-local NVMe storage or parallel file system. We enforce
this wait in order to avoid running out of the host memory since GPU-to-host copies are faster than host copies to persistent storage.

\paragraph{\bf Streamlined Multi-level Flushing to Persistent Storage:}
Although we coalesce the shards into a single pre-allocated memory region on the host memory, it is important
to note that it is not necessary to wait until all shards are successfully copied to the host memory before
starting the flushes to persistent storage. Instead, we can imagine a streaming pattern: as soon as partial
checkpointing data is copied from the GPU to the host memory, we can immediately flush it to the persistent
storage. Using this approach, two separate physical links (GPU-to-host and host-to-persistent storage) can be 
used in parallel to transfer the checkpointing data, which reduces the I/O overheads associated with
checkpointing. Furthermore, it is important to note that GPUs have a separate GPU-to-host hardware copy
engine. Therefore, the memory accesses on a GPU issued during the forward pass and the backward pass, 
regardless of whether to run computational kernels or to communicate
with other remote GPUs (through NVLinks and/or GPUDirect RDMA~\cite{potluri2013efficient}),
do not compete with the copies of the shards. Likewise, flushing from host memory to persistent storage uses an entirely different I/O path that does not interfere with the GPUs. As a consequence, our approach maximizes the use of the I/O paths needed for checkpointing, 
while it maximizes 
the overlapping with the training iterations without slowing them down due to interference 
or contention for shared resources. Thanks to this approach, except for unavoidable waits not sufficiently 
postponed by lazy non-blocking copies, training iterations can 
effectively progress almost undisturbed by checkpointing.

\paragraph{\bf Asynchronous Distributed Consolidation of Model and Optimizer Shards:}
While often overlooked, a significant source of overhead in the case of synchronous checkpointing 
is the consensus needed among the GPUs to validate all shards as being successfully saved to the
persistent storage. Only then can a global checkpoint be declared to hold a valid model parameter 
and optimizer state that can be later reused to restart the training or study its evolution.
Thanks to our asynchronous streamlined multi-level flushing, there is an opportunity to 
hide the consensus overhead: once each GPU finished flushing the shards to persistent storage,
it can enter into a consensus protocol asynchronously, which can perfectly overlap with the training 
iterations. Furthermore, it is possible to reduce the number of participants in the consensus
by introducing a hierarchic consolidation protocol that first validates the shards belonging to the 
same GPU, then the partition of shards belonging to the GPUs sharing the same compute node, and 
finally, all partitions belonging to all compute nodes. In this work, we have
considered a simple two-phase commit protocol, but we note that our approach is generic and can
accommodate more advanced consensus protocols that are tolerant to byzantine failures (e.g. Paxos,
Raft~\cite{howard2020paxos}).

\subsection{\proj Architecture}\label{sec:arch}

\begin{figure}[t]
    \centering
    \includegraphics[width=0.98\linewidth]{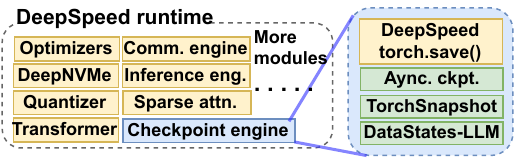}
    \caption{Three checkpointing engines added in DeepSpeed runtime (highlighted in green) for comparative evaluation.}
    \label{fig:engine-changes}
    \Description{
    
    }
\end{figure}

We implement our multi-level asynchronous checkpointing approach as a modular extension to the DeepSpeed runtime in the form of a checkpointing engine, as an alternative to the default synchronous engine (based on \verb|torch.save()|) and the asynchronous Nebula engine (which is closed-sources and exclusively available on Microsoft Azure cloud). This is illustrated in Figure~\ref{fig:engine-changes}. Our engine can be enabled in the configuration file which is supplied to the DeepSpeed engine at runtime and consists of a single attribute object specifying the size of the host buffer which can be reserved per process for caching checkpoints. Note that this extension does not utilize anything specific to DeepSpeed and can be easily adopted by other training runtimes as well.

We note that all checkpointing primitives and APIs of \proj
are the same as those used by DeepSpeed's default checkpointing engine, except for one additional method which blocks 
as long as any previous snapshot capture operations are pending. 
At the application level, checkpointing is transparent to the user, and no code modifications 
are needed to select any of the available checkpointing approaches, 
including the one that is proposed in \proj.
The integration of \proj was performed through DeepSpeed's fork of Megatron-LM, which contains ZeRO-based optimizations for the Megatron framework and does not need any modifications to use our checkpointing approach.

\subsection{Implementation}
\label{sec:impl}

Our checkpointing engine\footnote{The source code of \proj is available at \url{https://github.com/DataStates/datastates-llm}.}, is written in 
C++/CUDA and is exposed to DeepSpeed through Python and C++ APIs. The pinned host buffer is managed through a simple lightweight circular buffer manager, considering the producer-consumer pattern described in the design principles (\S~\ref{sec:design}). Dedicated CUDA streams and threads are used for device-to-host and host-to-file transfers. Such offloading of transfers and flushes in C++ enables our approach to overcome the limitations of the state-of-the-art asynchronous approaches (e.g., CheckFreq~\cite{mohanCheckFreqFrequentFineGrained}, LightCheck~\cite{lightCheck}, and Lightning's AsyncCheckpointIO~\cite{WelcomePyTorchLightning}) 
which perform background checkpointing and flushes through Python threads. These Python thread-based implementations are prone to inefficiencies arising from Python Global Interpreter Lock (GIL), lack of stream-based copies through GPU-copy engines supporting DMA, and host buffer re-allocation overheads.

Given a Python object (composed of tensors on both GPU and host memory, arrays, objects, and other data structures) that needs to be checkpointed, our checkpointing engine decomposes this operation into three phases as follows:
(1) recursively parse the Python object, and create a list of large arrays and tensors (across both GPU and host memory) by storing their memory pointers and sizes; (2) create a header by computing the file offsets for each tensor/object marked for asynchronous transfer in step (1); and (3) enqueue asynchronous device-to-host transfer (if required) and host-to-disk writes of headers, tensors and large objects (obtained in step-1).

\section{Performance Evaluation}
\label{sec:evaluation}

\subsection{Experimental Setup}
\label{sec:setup}

\paragraph{\bf Platform:} 
We conduct our experiments on ALCF's Polaris~\footnote{https://www.alcf.anl.gov/polaris}
HPC testbed. It consists of 560 nodes, each equipped with 512~GB of DDR4 memory~(aggregated from four NUMA domains), a 32-core AMD Zen 3 (Milan)~(64 threads), two 1.6 TB SSDs~(2~GB/s) and four Nvidia A100 GPUs aggregating to a total of 160~GB HBM memory. On each node, the four A100 GPUs are connected with each other using four NVLinks and with the host memory through a PCIe Gen 4 interface. The peak unidirectional Device-to-Device~(D2D), and pinned Device-to-Host~(D2H) (and vice versa) bandwidths on each GPU are 85~GB/s and 25~GB/s, respectively. There is a one-to-one mapping between the GPU and the NUMA domains, therefore concurrent device-to-host access by multiple GPUs does not create contention on the PCIe interface. 
The checkpoints are flushed to persistent storage, which is a Lustre~\cite{schwan2003lustre} parallel file system, 
composed of 160 Object Storage Targets~(OSTs) and 40 Metadata targets, with an aggregated bandwidth of 650~GB/s.

\paragraph{\bf Software:}
All the nodes run Nvidia CUDA driver 470.103, NVCC v11.8.89, Python v3.10, PyTorch v2.1, and DeepSpeed v0.11.2 on top of the Cray SUSE Linux Enterprise Server 15 operating system. In our experiments, we use up to 128 nodes (512 GPUs) to study the impact of large model sizes through data, tensor and pipeline parallelism, and contention of checkpoint flushes for the parallel file system.

\subsection{Compared Approaches}

\paragraph{\bf DeepSpeed:} 
This is the default checkpointing approach 
used in
the DeepSpeed~\cite{rasleyDeepSpeedSystemOptimizations2020}
LLM training runtime using PyTorch's default \verb|torch.save()| approach. This approach blocks the LLM training and performs synchronous writes of the checkpoint to the persistence storage, thereby providing consistency guarantees for the checkpoint (illustrated as \emph{(a) DeepSpeed default synchronous checkpointing} in Figure~\ref{fig:overlap-llm-flow}).

\paragraph{\bf Asynchronous Checkpointing:} 
This approach is representative of the in-memory snapshotting techniques adopted by CheckFreq~\cite{mohanCheckFreqFrequentFineGrained}, LightCheck\cite{lightCheck}, and PyTorch Lightning's AsyncCheckpointIO~\cite{WelcomePyTorchLightning} (illustrated as \emph{(b) Asynchronous checkpointing} in Figure~\ref{fig:overlap-llm-flow}), and is replicated to mimic AsyncCheckpointIO~\cite{LightningAsyncCheckpointIO} (we had
to adapt such techniques for LLMs since the original implementations do not support pipeline and tensor parallelism). Specifically, in the first phase, it allocates
a buffer for each shard on the host memory (red block), then copies the shard
from the device to the host buffer (green blocks). Once the first phase has
finished, it proceeds to asynchronously flush the shards from the host memory
to persistent storage (Lustre PFS in our case). This is depicted in Figure~\ref{fig:overlap-llm-flow}(b). The allocation overhead can be significant
due to the need to pin the host memory~\cite{VELOCGPU-HiPC22}, especially when considering a large
number of shards. It highlights an important limitation of many state-of-the-art approaches that are not optimized for LLM checkpointing. 

\paragraph{\bf TorchSnapshot:}
This is a state-of-the-art checkpointing runtime developed by the PyTorch team  (illustrated as \emph{(c) TorchSnapshot} in Figure~\ref{fig:overlap-llm-flow}). It optimizes checkpointing by (1) parallelizing state capture across data-parallel replicas (which is moot for DeepSpeed/Megatron since the latter shards the checkpoints by default); (2) splitting tensors in chunks for overlapping transfers in streaming fashion from the device-to-host and host-to-disk; and (3) multi-threaded write of chunked tensors in different files on the disk, thereby utilizing higher disk write bandwidth, but incurring additional metadata and flushing overheads because of larger number of files~\cite{IOAgg-ISPDC23}. We limit the number of parallel flush threads per GPU to 4, which shows peak write throughput to persistent storage in our experimental testbed.

\paragraph{\bf \proj (Our Approach):}
This is the implementation of \proj based on the 
design proposed in \S~\ref{sec:designandimpl} and illustrated as \emph{(d) \proj} in Figure~\ref{fig:overlap-llm-flow}.

\subsection{Evaluation Methodology}

\begin{table}
    \centering
    \caption{Configuration of models and runtime used for evaluations derived from BLOOM~\cite{workshopBLOOM176BParameterOpenAccess2023} (highlighted by gray column) and LLaMA~\cite{touvronLlamaOpenFoundation2023}.}
    \label{tab:models}
    \setlength{\tabcolsep}{5pt}
    \begin{tabular}{|c||>{\columncolor[gray]{0.8}}c|c|c|c|c|}
    \hline
    Model size in billions & 3 & 7 & 13 & 30 & 70 \\
    \hline
    \hline
    Layers & 30 & 32 & 40 & 60 & 80 \\
    Hidden dim. & 2560 & 4096 & 5120 & 6656 & 8192 \\
    Atten. heads & 32 & 32 & 40 & 52 & 64 \\
    Num. of nodes & 1 & 2 & 4 & 8 & 20  \\
    Tensor parallelism & \multicolumn{5}{|c|}{4 (=Number of GPUs per node)} \\
    Pipeline parallelism & \multicolumn{5}{|c|}{=Number of nodes} \\
    ZeRO optimization & \multicolumn{5}{|c|}{Stage 1 (Partition optimizer state)} \\   
    \hline
    \end{tabular}
\end{table}

\paragraph{\bf Models, Sharding, and Dataset:}
We use five different LLM model sizes in our evaluations based on the real-world setups: BLOOM (3B)~\cite{workshopBLOOM176BParameterOpenAccess2023}, LLaMA (30B), and LLaMA2 (7B, 13B, 70B)~\cite{touvronLlamaOpenFoundation2023} model architectures. The models and their runtime configurations are summarized in Table~\ref{tab:models}.

To minimize the intra-layer communication overheads, the tensor-parallel degree is set to 4, 
which is the number of GPUs in a single node and all are interconnected through fast NVLinks. 
To fit the model across distributed GPU memories, the pipelines are split evenly across the number of nodes described in Table~\ref{tab:models} using the default partitioning scheme of uniformly balancing the number of trainable parameters on each pipeline stage. Unless otherwise noted, the data-parallelism degree is set to 1, representing a single LLM replica being used for training. For the experiments that
involve the data parallelism approach, the optimizer state is sharded across the replicas.
This corresponds to the configuration Figure~\ref{fig:ckpt-sharding}(d).

Throughout our experiments, we use a subset of the OSCAR-en dataset 
included in the repository of the BLOOM model. It consists of 79K records, \cite{workshopBLOOM176BParameterOpenAccess2023}, and use the default LLaMA2 \cite{touvronLlamaOpenFoundation2023} tokenizer for pre-processing the dataset into tokens. Similar to BLOOM training, the default sequence length is set to 2048, and the micro-batch size is 16 to avoid out-of-memory (OOM) errors in any configuration.

\paragraph{\bf Memory and Storage Tiers:}
Each of the compared approaches is allowed to use up to a maximum of
64~GB of host memory, the rest of which is reserved for caching the training
data.  Since the average checkpoint size per GPU is 10-15~GB (shown in Figure~\ref{fig:agg-ckpt-size}) and there are four GPUs per node, this is 
enough to hold a full checkpoint across all compute nodes. From the host
memory, the checkpoint shards are flushed directly to Lustre, which
acts as the shared persistent storage.

\paragraph{\bf Key Performance Metrics:}
Throughout our evaluations, we measure the following metrics for comparing the aforementioned approaches: 
(1) checkpointing throughput of different model sizes to evaluates the blocking checkpointing overhead on the application for a broad range of increasing complex LLMs; 
(2) impact on iteration duration during checkpointing to evaluate the slowdown and interference caused by checkpointing on training iterations; and 
(3) end-to-end training runtime to study the broader impact on overall job completion times.
We evaluate the above metrics under different settings: 
(a) varying degrees of data parallelism since DeepSpeed runtime partitions the checkpoints across data-parallel ranks for faster checkpointing, this setting studies the impact of strong scaling (more flushing bandwidth available to capture the checkpoint of the same size), and 
(b) varying checkpointing frequency to study how the training performs for different degrees of I/O pressure arising from frequent or sparse checkpointing scenarios.

\subsection{Performance Results}

\begin{figure*}[t]
\centering
\minipage{0.49\textwidth}
    \centering
    \includegraphics[width=\linewidth]{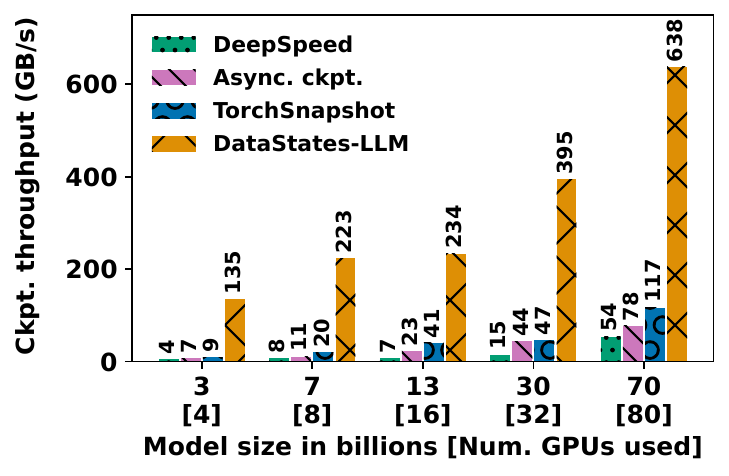}
    \caption{Aggregate checkpointing throughput for different model sizes. Higher is better.}
    \label{fig:ckpt-thru-diff-model-sizes}
\endminipage\hfill
\minipage{0.49\textwidth}
    \centering
    \includegraphics[width=\linewidth]{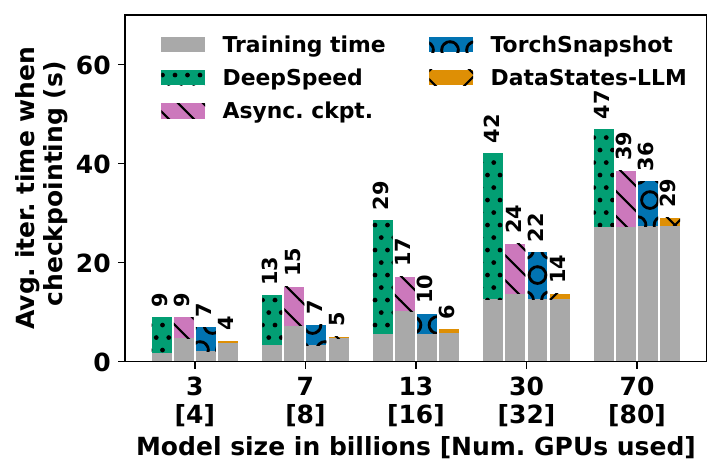}
    \caption{Average training iteration time for different model sizes when checkpointing. Lower is better.}
    \label{fig:per-process-iter-diff-model-sizes}
\endminipage\hfill
\minipage{0.49\textwidth}
    \centering
    \includegraphics[width=\linewidth]{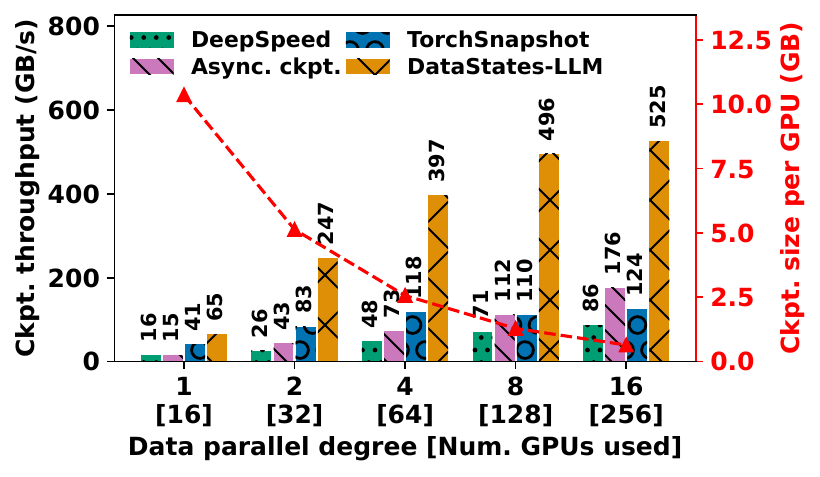}
    \caption{Aggregate checkpointing throughput for a 13B model for different data-parallel degrees. Higher is better.}
    \label{fig:scale-dp-13B}
\endminipage\hfill
\minipage{0.49\textwidth}
    \centering
    \includegraphics[width=\linewidth]{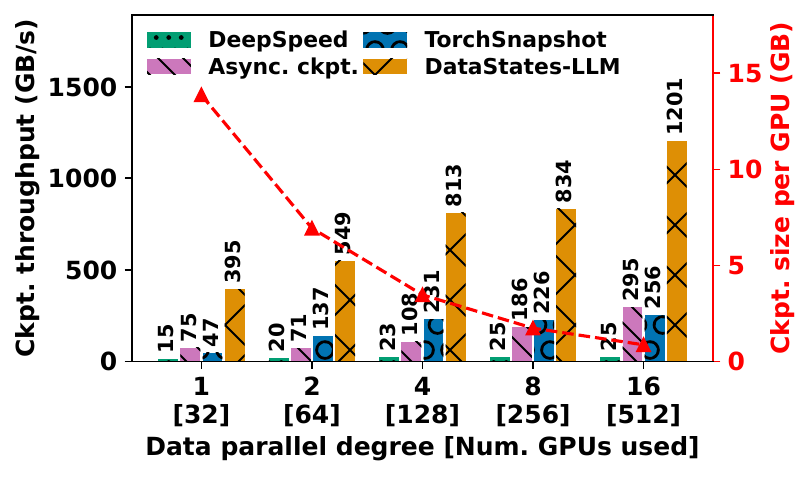}
    \caption{Aggregate checkpointing throughput for a 30B model for different data-parallel degrees. Higher is better.}
    \label{fig:scale-dp-30B}
\endminipage\hfill
\Description{A set of figures illustrating aggregate checkpointing throughput 
for different model sizes, average training iteration time for different model sizes when checkpointing, aggregate checkpointing throughput for a 13B model for different data-parallel degrees, aggregate checkpointing throughput for a 30B model for different data-parallel degrees.}
\end{figure*}

\paragraph{\bf Increasing LLM Model Size Without Data Parallelism:}

In our first set of experiments, we evaluate the following two metrics for increasing model sizes: (1) the average checkpointing throughput perceived by the training process, which is defined as the total checkpoint size divided by the time for which the training was blocked for each checkpointing operation; and (2) the average iteration duration when checkpointing, which shows the overheads of checkpointing on the training process in both direct form --- the amount of time for which training is blocked to capture checkpoint, and indirect form --- slowdown in training process caused by interference from checkpointing I/O. The training is run for five iterations with a checkpoint being taken at every iteration. Such high-frequency checkpointing at every iteration allows us to study the performance overheads of different approaches under high I/O pressure. We note two interesting observations for evaluating this metric. First, Since the asynchronous checkpoint operations from device-to-host and host-to-file overlap with the computations of the next iterations, from an application perspective, this metric is important to study the checkpointing stalls experienced by the application by different checkpointing approaches. Second, the checkpoint operation is a blocking collective with respect to the model and optimizer update stage during training, i.e., none of the processes can start updating the model or optimizer states until all parts of the previous checkpoint are consistently captured either on the host memory or on the persistent file. Therefore, the checkpointing throughput observed by the application is dictated by the slowest process across all processes. 

As observed in Figure~\ref{fig:ckpt-thru-diff-model-sizes}, the checkpointing throughput increases with increasing model size. This is because of two reasons: (1) The training duration per iteration increases with larger models due to the higher complexity of transformer layers and higher communication overheads (for sharing activations, gradients, optimizer partitions, and model updates) across multiple nodes (as depicted in Figure~\ref{fig:diff-phases-time}). The increasing iteration duration allows for more time to asynchronously flush the previous checkpoints, thereby not blocking future checkpoint requests due to pending flushes. (2) Larger models are run on more number of nodes (as outlined in Table~\ref{tab:models}), leading to more device-to-host interconnects which can be exploited for parallel flushing of checkpoints between node-local memory tiers, and higher write bandwidth available for flushing checkpoints to the persistent file system. As a consequence of the above two factors, we observe a linear scalability trend of checkpointing throughput in Figure~\ref{fig:ckpt-thru-diff-model-sizes} for all approaches. However, compared to DeepSpeed, Asynchronous checkpointing, or TorchSnapshot, \proj demonstrates at least 4$\times$ and up to 34$\times$ higher checkpointing throughput across various model sizes. 

Next, we study the impact on the overall iteration duration. Figure~\ref{fig:per-process-iter-diff-model-sizes} shows the breakdown of per-process iteration duration as training time vs. checkpointing time. We observe that the training time (consisting of forward pass, backward pass, and update phases) of smaller models (3B, 7B, 13B, and 30B) are similar for all approaches except for the Asynchronous checkpointing approach. This is because of the interference caused by slow host-memory allocation, slow transfers to unpinned host-memory, and PCIe contention with loading the next micro-batch on the GPU from the data pipeline. This effect is not observed in the larger 70B model because, for large models with the same amount of checkpoint data per GPU (shown in Figure~\ref{fig:agg-ckpt-size}), the long forward and backward passes amortize the slow allocation and transfer overheads. With increasing model size, the training time increases (Figure~\ref{fig:diff-phases-time}), while the checkpoint size per GPU remains consistent (Figure~\ref{fig:agg-ckpt-size}). Therefore, the ratio of the training 
duration to blocking duration while waiting for checkpoints to finish increases with the model size. However, irrespective of the fact that the training phase dictates the major proportion of the iteration time, \proj speeds up the iteration by at least 23\%, and up to 4.5$\times$ compared to other approaches we studied in evaluating \proj.

\paragraph{\bf Fixed LLM Model Size with Increasing Data Parallelism:}

In our next set of experiments, we evaluate the checkpointing throughput as a function of increasing degrees of data parallelism. Similar to the previous set of experiments, we conducted this experiment by checkpointing during each of five consecutive iterations. This evaluation is important to study the efficiency of concurrent flushing of the partitioned optimizer state across the data parallel replicas. We evaluate the checkpointing throughput by scaling the data parallelism degrees from 1 to 16 for two model sizes: 13B and 30B. We do not consider the smaller 3B and 7B models because at high degrees of data parallelism, such models are partitioned at excessive levels, which results in tiny 
shards that lead to the underutilization of GPUs. On the other hand, large models such as 70B show similar trends as the 30B model, but run for much longer.
We only scale up to a data-parallel degree of 16 with 512 GPUs because it is not trivial to train a large number of data-parallel replicas in practice due to the high costs of GPU resources --- for instance, BLOOM 175B was trained with 8 data-parallel replicas on a total of 384 GPUs.

\begin{figure*}[ht]
\centering
\begin{subfigure}[h]{0.32\linewidth}
    \centering
    \includegraphics[width=\linewidth]{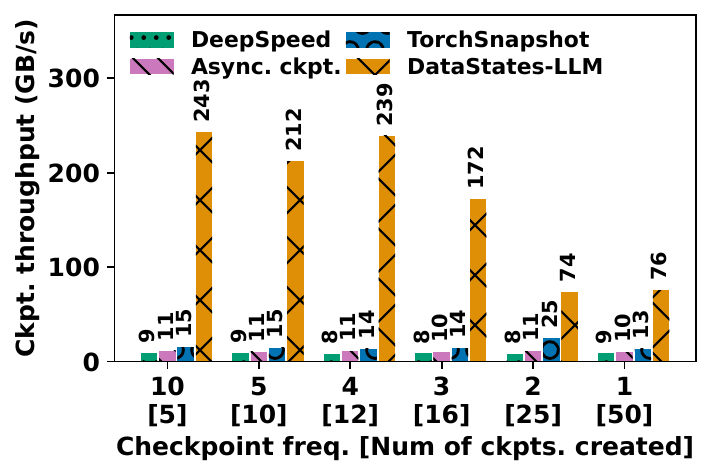}
    \caption{Aggregate checkpointing throughput. Higher is better.}
    \label{fig:vary-ckpt-iter-ckpt-thru-7B}
\end{subfigure}
\hfill
\begin{subfigure}[h]{0.32\linewidth}
    \centering
    \includegraphics[width=\linewidth]{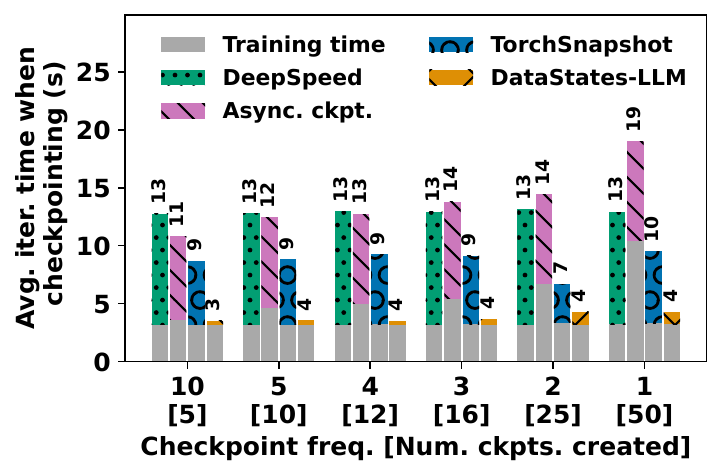}
    \caption{Per process iteration time when checkpointing. Lower is better.}
    \label{fig:vary-ckpt-iter-iter-time-7B}
\end{subfigure}
\hfill
\begin{subfigure}[h]{0.32\linewidth}
    \centering
    \includegraphics[width=\linewidth]{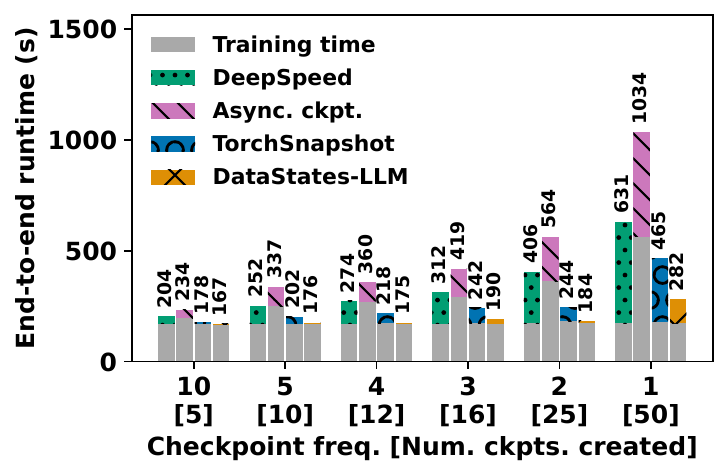}
    \caption{End-to-end training time. Lower is better.}
    \label{fig:vary-ckpt-iter-e2e-7B}
\end{subfigure}
\caption{Running training for 50 iterations for a 7B model with different checkpointing frequencies.}
\Description{A set of figures illustrating aggregate checkpointing 
throughput, per process iteration duration when checkpointing,
and end-to-end training time for different checkpointing frequencies
when training a 7B model for a total of 50 iterations.}
\label{fig:scale-ckpt-freq-7B}
\end{figure*}

\begin{figure*}[ht]
\begin{subfigure}[t]{0.32\linewidth}
    \includegraphics[width=\linewidth]{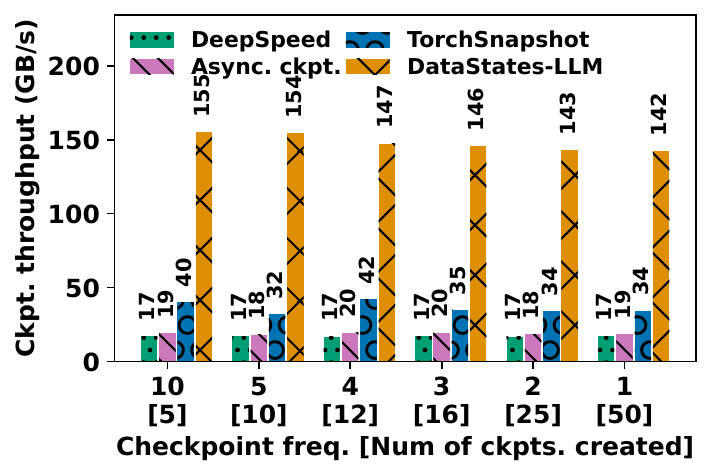}
    \caption{Aggregate checkpointing throughput. Higher is better.}
    \label{fig:vary-ckpt-iter-ckpt-thru-13B}
\end{subfigure}
\hfill
\begin{subfigure}[t]{0.32\linewidth}
    \includegraphics[width=\linewidth]{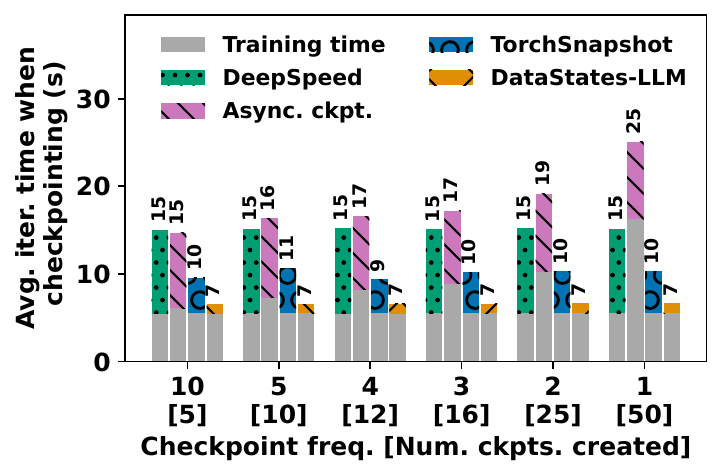}
    \caption{Per process iteration time when checkpointing. Lower is better.}
    \label{fig:vary-ckpt-iter-iter-time-13B}
\end{subfigure}
\hfill
\begin{subfigure}[t]{0.32\linewidth}
    \includegraphics[width=\linewidth]{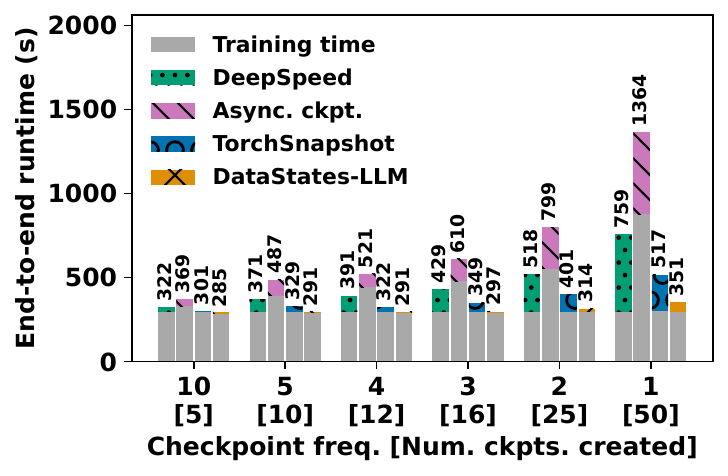}
    \caption{End-to-end training time. Lower is better.}
    \label{fig:vary-ckpt-iter-e2e-13B}
\end{subfigure}
\caption{Running training for 50 iterations for a 13B model with different checkpointing frequencies.}
\label{fig:scale-ckpt-freq-13B}
\Description{A set of figures illustrating aggregate checkpointing 
throughput, per process iteration duration when checkpointing,
and end-to-end training time for different checkpointing frequencies
when training a 13B model for a total of 50 iterations.}
\end{figure*}

Figure~\ref{fig:scale-dp-13B} and Figure~\ref{fig:scale-dp-30B} show the checkpointing throughput with increasing scale of data parallelism for the 13B and 30B models. We observe that the checkpoint size per GPU, referenced by dashed-red lines on the minor y-axis, shows a linear decrease of checkpoint size per GPU with increasing degrees of data parallel replicas. Therefore, this study captures the strong scalability of checkpoint performance, i.e., how well can various checkpointing approaches perform when the same checkpoint is distributed across multiple ranks, such that they can be flushed in parallel. More specifically, the checkpoint size per GPU drops from $\sim$10.4~GB to $\sim$650~MB per GPU for the 13B model, and from $\sim$13.8~GB to $\sim$870~MB per GPU for the 30B model, when scaling the data parallel degree from 1 to 16. When comparing the 13B and 30B models for the same number of GPUs (e.g., the 13B model with DP=4 and 30B model with DP=2 for 64 GPUs), we see that the checkpointing throughput of the 13B model is lower than the 30B model even though both approaches have the same number of parallel channels for flushing the checkpoint. This is because the training iteration of the 13B model is significantly faster than the 30B model and therefore needs to stall training for checkpointing more frequently as compared to the long-running iteration of the 30B model. While all approaches scale well to the increasing data parallel replicas due to concurrent flushes, our approach outperforms the DeepSpeed synchronous, Asynchronous checkpointing approach, and TorchSnapshot by 2.8$\times$, 1.75$\times$, and 1.78$\times$, respectively for the 13B model; and for the 30B model by 48$\times$, 4.12$\times$, and 4.7$\times$, respectively. In terms of end-to-end training runtime of the 30B model, we observe that \proj shows up to 2.5$\times$ to 1.86$\times$ faster training completion time when scaling from DP=1 to DP=16 as compared to other approaches. Similar trends are observed for the 13B model. Therefore, our approach excels at strong scalability experiments of checkpointing and demonstrates significant speedup in end-to-end training runtimes.

\paragraph{\bf Increasing Checkpointing Frequency:}

Next, we study the impact of scaling the checkpoint frequency, i.e., the number of iterations elapsed between consecutive checkpoint operations. This allows us to understand the efficiency of overlapping between the training and asynchronous checkpoint flushes such that the large intervals between subsequent checkpoint operations would allow for more time to complete the flushes to persistent storage and free up the host-memory buffer for the next checkpoints. 

In particular, we evaluate the checkpointing throughput, iteration slowdown caused due to checkpointing, and the end-to-end runtime for a variable number of checkpoints captured during a 50-iteration run of the 7B and 13B models.
Thanks to fast forward and backward passes, the 7B model presents less
opportunities to overlap asynchronous I/O with the training iterations.
Therefore, we chose it to highlight the difference between the approaches
when the I/O pressure dominates. Conversely, the 13B model captures 
the opposite trend observed in larger model, where slower forward and 
backward passes enable more opportunities for overlap.

For the 7B model, we observe in Figure~\ref{fig:vary-ckpt-iter-ckpt-thru-7B} that the checkpointing throughput of \proj decreases with an increasing checkpointing frequency due to higher I/O pressure, which arises due to the bottleneck of slow checkpoint flushes to the disk. On the other hand, the 13B model, depicted in Figure~\ref{fig:vary-ckpt-iter-ckpt-thru-13B}, does
not exhibit this effect. Instead, the checkpointing throughput remains
high regardless of the checkpointing frequency. In any case, the 
other approaches suffer from I/O bottlenecks regardless of model size.
As a consequence, \proj achieves at least 3$\times$ higher checkpointing throughput for the 7B model and 4.2$\times$ higher checkpointing throughput
for the 13B model.

Furthermore, we observe in Figure~\ref{fig:vary-ckpt-iter-iter-time-7B} and Figure~\ref{fig:vary-ckpt-iter-iter-time-13B}, respectively, that with increasing checkpointing frequency, the Asynchronous checkpointing approach slows down the training phase significantly, due to slow host memory allocation and transfers, similar to the effect illustrated in Figure~\ref{fig:per-process-iter-diff-model-sizes}. On the other hand, the other compared approaches do not
increase the duration of the training iteration. However, thasnks to better
overlapping with the forward and backward pass, \proj achieves at least 1.3$\times$ and up to 3.8$\times$ faster iteration duration during checkpointing as compared with the other approaches. 

Lastly, we study the end-to-end time taken to complete the entire training process, including the pending flushes towards the end of training. Figure~\ref{fig:vary-ckpt-iter-e2e-7B} and Figure~\ref{fig:vary-ckpt-iter-e2e-13B} depict the end-to-end runtime of the 7B model and the 13B model, respectively. The end-to-end training runtime shows performance trends similar to those observed in iteration-scale analysis (Figure~\ref{fig:vary-ckpt-iter-iter-time-7B} and Figure~\ref{fig:vary-ckpt-iter-iter-time-13B}).
Specifically, our approach remains up to 3.86$\times$ faster in end-to-end training as compared to the other approaches even for an increasing checkpointing frequency. 

\section{Conclusions}
In this work, we address the problem of high overheads incurred due to checkpointing in large-scale distributed LLM training running with advanced hybrid parallelism strategies using widely adopted runtimes such as DeepSpeed. State-of-the-art checkpoint engines, specifically designed for LLMs slow down the training while checkpointing because (1) they do not exploit the characteristics of various training phases to overlap checkpoint I/O efficiently; and (2) they underutilize the available interconnects and memory resources, leading to significant stalls during training. The checkpointing overheads are exacerbated when model and/or optimizer states need to be frequently checkpointed for defensive and productive use cases. To address these limitations, we design and develop \proj, which efficiently and transparently overlaps the checkpoint I/O with the \emph{immutable} phases of forward and backward passes during training. \proj proposes key design ideas to mitigate checkpoint overheads in LLMs, such as preallocating and reusing pinned host buffer for fast DMA transfers, coalescing of model/optimizer shards while transferring checkpoints from GPU to host-memory, lazy non-blocking checkpoint snapshotting overlapping with forward and backward training phases, streaming multi-level flushing to persistent storage, and asynchronous distributed consensus of checkpoint persistence. We ran extensive evaluations with varying model sizes derived from production-grade runs of BLOOM and LLaMA2, different data parallelism configurations, and checkpointing frequency intervals.
Results show that \proj checkpoints 3$\times$ to 4.2$\times$ faster than existing state-of-the-art checkpointing runtimes, which achieves a speedup of the end-to-end training by 1.3$\times$ to 2.2$\times$.

Encouraged by these promising results, in future 
we plan to explore data reduction techniques such as differential checkpointing and compression to further minimize the network and storage costs when checkpointing at high frequencies. Furthermore, we will explore efficient checkpointing strategies 
when model and/or optimizer states are offloaded across multiple memory tiers. Finally, we did not study the metadata overheads resulting from storing each
shard as a separate file. This may lead to interesting trade-offs that justify investigating
novel aggregation and consolidation strategies.

\section*{Acknowledgements}
This work is supported in part by the U.S. Department of Energy (DOE), Office of Science, Office of Advanced Scientific Computing Research under contract DEAC02-06CH11357/0F-60169 and the National Science Foundation (NSF) under award no.\ 2106634/2106635. Results presented in this paper are obtained using Argonne's ALCF HPC systems, and NSF Cloudlab and Chameleon testbeds.

\balance
\bibliographystyle{ACM-Reference-Format}
\bibliography{references}


\begin{thebibliography}{46}


\ifx \showCODEN    \undefined \def \showCODEN     #1{\unskip}     \fi
\ifx \showDOI      \undefined \def \showDOI       #1{#1}\fi
\ifx \showISBNx    \undefined \def \showISBNx     #1{\unskip}     \fi
\ifx \showISBNxiii \undefined \def \showISBNxiii  #1{\unskip}     \fi
\ifx \showISSN     \undefined \def \showISSN      #1{\unskip}     \fi
\ifx \showLCCN     \undefined \def \showLCCN      #1{\unskip}     \fi
\ifx \shownote     \undefined \def \shownote      #1{#1}          \fi
\ifx \showarticletitle \undefined \def \showarticletitle #1{#1}   \fi
\ifx \showURL      \undefined \def \showURL       {\relax}        \fi
\providecommand\bibfield[2]{#2}
\providecommand\bibinfo[2]{#2}
\providecommand\natexlab[1]{#1}
\providecommand\showeprint[2][]{arXiv:#2}

\bibitem[Ansel et~al\mbox{.}(2009)]%
        {ansel2009dmtcp}
\bibfield{author}{\bibinfo{person}{Jason Ansel}, \bibinfo{person}{Kapil Arya}, {and} \bibinfo{person}{Gene Cooperman}.} \bibinfo{year}{2009}\natexlab{}.
\newblock \showarticletitle{DMTCP: Transparent checkpointing for cluster computations and the desktop}. In \bibinfo{booktitle}{\emph{IPDPS'09: International Symposium on Parallel \& Distributed Processing}}. \bibinfo{publisher}{IEEE}, \bibinfo{address}{Rome, Italy}, \bibinfo{pages}{1--12}.
\newblock


\bibitem[Arif et~al\mbox{.}(2022)]%
        {canary-sc22}
\bibfield{author}{\bibinfo{person}{Moiz Arif}, \bibinfo{person}{Kevin Assogba}, {and} \bibinfo{person}{M.~Mustafa Rafique}.} \bibinfo{year}{2022}\natexlab{}.
\newblock \showarticletitle{Canary: Fault-Tolerant FaaS for Stateful Time-Sensitive Applications}. In \bibinfo{booktitle}{\emph{SC22: International Conference for High Performance Computing, Networking, Storage and Analysis}}. \bibinfo{publisher}{IEEE}, \bibinfo{address}{Dallas, TX, USA}, \bibinfo{pages}{1--16}.
\newblock


\bibitem[Bautista-Gomez et~al\mbox{.}(2011)]%
        {bautista2011fti}
\bibfield{author}{\bibinfo{person}{Leonardo Bautista-Gomez}, \bibinfo{person}{Seiji Tsuboi}, \bibinfo{person}{Dimitri Komatitsch}, \bibinfo{person}{Franck Cappello}, \bibinfo{person}{Naoya Maruyama}, {and} \bibinfo{person}{Satoshi Matsuoka}.} \bibinfo{year}{2011}\natexlab{}.
\newblock \showarticletitle{FTI: High performance Fault Tolerance Interface for hybrid systems}. In \bibinfo{booktitle}{\emph{SC'11: Proceedings of the International Conference for High Performance Computing, Networking, Storage and Analysis}}. \bibinfo{publisher}{IEEE}, \bibinfo{address}{Seattle, WA, USA}, \bibinfo{pages}{1--12}.
\newblock


\bibitem[Chen et~al\mbox{.}(2023)]%
        {lightCheck}
\bibfield{author}{\bibinfo{person}{Menglei Chen}, \bibinfo{person}{Yu Hua}, \bibinfo{person}{Rong Bai}, {and} \bibinfo{person}{Jianming Huang}.} \bibinfo{year}{2023}\natexlab{}.
\newblock \showarticletitle{A Cost-Efficient Failure-Tolerant Scheme for Distributed DNN Training}. In \bibinfo{booktitle}{\emph{ICCD'23: Proceedings of the International Conference on Computer Design}}. \bibinfo{publisher}{IEEE}, \bibinfo{address}{Milan, Italy}, \bibinfo{pages}{150--157}.
\newblock


\bibitem[Choquette et~al\mbox{.}(2021)]%
        {a100-nvlink}
\bibfield{author}{\bibinfo{person}{Jack Choquette}, \bibinfo{person}{Wishwesh Gandhi}, \bibinfo{person}{Olivier Giroux}, \bibinfo{person}{Nick Stam}, {and} \bibinfo{person}{Ronny Krashinsky}.} \bibinfo{year}{2021}\natexlab{}.
\newblock \showarticletitle{NVIDIA A100 Tensor Core GPU: Performance and Innovation}.
\newblock \bibinfo{journal}{\emph{IEEE Micro}} \bibinfo{volume}{41}, \bibinfo{number}{2} (\bibinfo{year}{2021}), \bibinfo{pages}{29--35}.
\newblock


\bibitem[Chowdhery et~al\mbox{.}(2023)]%
        {chowdhery2023palm}
\bibfield{author}{\bibinfo{person}{Aakanksha Chowdhery}, \bibinfo{person}{Sharan Narang}, \bibinfo{person}{Jacob Devlin}, \bibinfo{person}{Maarten Bosma}, \bibinfo{person}{Gaurav Mishra}, \bibinfo{person}{Adam Roberts}, \bibinfo{person}{Paul Barham}, \bibinfo{person}{Hyung~Won Chung}, \bibinfo{person}{Charles Sutton}, \bibinfo{person}{Sebastian Gehrmann}, {et~al\mbox{.}}} \bibinfo{year}{2023}\natexlab{}.
\newblock \showarticletitle{Palm: Scaling language modeling with pathways}.
\newblock \bibinfo{journal}{\emph{JMLR'23: Journal of Machine Learning Research}} \bibinfo{volume}{24}, \bibinfo{number}{240} (\bibinfo{year}{2023}), \bibinfo{pages}{1--113}.
\newblock


\bibitem[Fedus et~al\mbox{.}(2022)]%
        {google-switch}
\bibfield{author}{\bibinfo{person}{William Fedus}, \bibinfo{person}{Barret Zoph}, {and} \bibinfo{person}{Noam Shazeer}.} \bibinfo{year}{2022}\natexlab{}.
\newblock \showarticletitle{Switch transformers: scaling to trillion parameter models with simple and efficient sparsity}.
\newblock \bibinfo{journal}{\emph{JMLR'22: Journal of Machine Learning Research}} \bibinfo{volume}{23}, \bibinfo{number}{1}, Article \bibinfo{articleno}{120} (\bibinfo{date}{jan} \bibinfo{year}{2022}), \bibinfo{numpages}{39}~pages.
\newblock
\showISSN{1532-4435}


\bibitem[Godoy et~al\mbox{.}(2020)]%
        {godoy2020adios}
\bibfield{author}{\bibinfo{person}{William~F Godoy}, \bibinfo{person}{Norbert Podhorszki}, \bibinfo{person}{Ruonan Wang}, \bibinfo{person}{Chuck Atkins}, \bibinfo{person}{Greg Eisenhauer}, \bibinfo{person}{Junmin Gu}, \bibinfo{person}{Philip Davis}, \bibinfo{person}{Jong Choi}, \bibinfo{person}{Kai Germaschewski}, \bibinfo{person}{Kevin Huck}, {et~al\mbox{.}}} \bibinfo{year}{2020}\natexlab{}.
\newblock \showarticletitle{Adios 2: The adaptable input output system. a framework for high-performance data management}.
\newblock \bibinfo{journal}{\emph{SoftwareX}}  \bibinfo{volume}{12} (\bibinfo{year}{2020}), \bibinfo{pages}{100561}.
\newblock


\bibitem[Gossman et~al\mbox{.}(2023)]%
        {IOAgg-ISPDC23}
\bibfield{author}{\bibinfo{person}{Mikaila Gossman}, \bibinfo{person}{Bogdan Nicolae}, {and} \bibinfo{person}{Jon Calhoun}.} \bibinfo{year}{2023}\natexlab{}.
\newblock \showarticletitle{Modeling Multi-Threaded Aggregated I/O for Asynchronous Checkpointing on HPC Systems}. In \bibinfo{booktitle}{\emph{ISPDC'23: Proceedings of the International Conference on Parallel and Distributed Computing}}. \bibinfo{publisher}{IEEE}, \bibinfo{address}{Bucharest, Romania}, \bibinfo{pages}{101--105}.
\newblock
\urldef\tempurl%
\url{https://hal.inria.fr/hal-04343661}
\showURL{%
\tempurl}


\bibitem[Hargrove and Duell(2006)]%
        {hargrove2006berkeley}
\bibfield{author}{\bibinfo{person}{Paul~H Hargrove} {and} \bibinfo{person}{Jason~C Duell}.} \bibinfo{year}{2006}\natexlab{}.
\newblock \showarticletitle{{Berkeley lab checkpoint/restart (blcr) for linux clusters}}.
\newblock \bibinfo{journal}{\emph{IOP Publishing}} \bibinfo{volume}{46}, \bibinfo{number}{1} (\bibinfo{year}{2006}), \bibinfo{pages}{494}.
\newblock


\bibitem[He et~al\mbox{.}(2016)]%
        {resnet}
\bibfield{author}{\bibinfo{person}{Kaiming He}, \bibinfo{person}{Xiangyu Zhang}, \bibinfo{person}{Shaoqing Ren}, {and} \bibinfo{person}{Jian Sun}.} \bibinfo{year}{2016}\natexlab{}.
\newblock \showarticletitle{Deep Residual Learning for Image Recognition}. In \bibinfo{booktitle}{\emph{Conference on Computer Vision and Pattern Recognition (CVPR)}}. \bibinfo{publisher}{IEEE}, \bibinfo{address}{Las Vegas, USA}, \bibinfo{pages}{770--778}.
\newblock


\bibitem[He et~al\mbox{.}(2023)]%
        {he2023unicron}
\bibfield{author}{\bibinfo{person}{Tao He}, \bibinfo{person}{Xue Li}, \bibinfo{person}{Zhibin Wang}, \bibinfo{person}{Kun Qian}, \bibinfo{person}{Jingbo Xu}, \bibinfo{person}{Wenyuan Yu}, {and} \bibinfo{person}{Jingren Zhou}.} \bibinfo{year}{2023}\natexlab{}.
\newblock \bibinfo{title}{Unicron: Economizing Self-Healing LLM Training at Scale}.
\newblock
\newblock
\showeprint[arxiv]{2401.00134}~[cs.DC]


\bibitem[Howard and Mortier(2020)]%
        {howard2020paxos}
\bibfield{author}{\bibinfo{person}{Heidi Howard} {and} \bibinfo{person}{Richard Mortier}.} \bibinfo{year}{2020}\natexlab{}.
\newblock \showarticletitle{Paxos vs Raft: have we reached consensus on distributed consensus?}. In \bibinfo{booktitle}{\emph{PaPoC'20: The 7th Workshop on Principles and Practice of Consistency for Distributed Data}}. \bibinfo{publisher}{ACM}, \bibinfo{address}{Heraklion, Greece}, Article \bibinfo{articleno}{8}, \bibinfo{numpages}{9}~pages.
\newblock
\showISBNx{9781450375245}


\bibitem[Huang et~al\mbox{.}(2019)]%
        {huang2019gpipe}
\bibfield{author}{\bibinfo{person}{Yanping Huang}, \bibinfo{person}{Youlong Cheng}, \bibinfo{person}{Ankur Bapna}, \bibinfo{person}{Orhan Firat}, \bibinfo{person}{Dehao Chen}, \bibinfo{person}{Mia Chen}, \bibinfo{person}{HyoukJoong Lee}, \bibinfo{person}{Jiquan Ngiam}, \bibinfo{person}{Quoc~V Le}, \bibinfo{person}{Yonghui Wu}, {and} \bibinfo{person}{zhifeng Chen}.} \bibinfo{year}{2019}\natexlab{}.
\newblock \showarticletitle{GPipe: Efficient Training of Giant Neural Networks using Pipeline Parallelism}. In \bibinfo{booktitle}{\emph{NeurIPS'19: Advances in Neural Information Processing Systems}}, \bibfield{editor}{\bibinfo{person}{H.~Wallach}, \bibinfo{person}{H.~Larochelle}, \bibinfo{person}{A.~Beygelzimer}, \bibinfo{person}{F.~d\textquotesingle Alch\'{e}-Buc}, \bibinfo{person}{E.~Fox}, {and} \bibinfo{person}{R.~Garnett}} (Eds.), Vol.~\bibinfo{volume}{32}. \bibinfo{publisher}{Curran Associates, Inc.}, \bibinfo{address}{Vancouver, Canada}.
\newblock


\bibitem[Kingma and Ba(2017)]%
        {kingma2014adam}
\bibfield{author}{\bibinfo{person}{Diederik~P. Kingma} {and} \bibinfo{person}{Jimmy Ba}.} \bibinfo{year}{2017}\natexlab{}.
\newblock \bibinfo{title}{Adam: A Method for Stochastic Optimization}.
\newblock
\newblock
\showeprint[arxiv]{1412.6980}~[cs.LG]


\bibitem[Li et~al\mbox{.}(2020)]%
        {DBLP:journals/pvldb/LiZVSNLPSVDC20}
\bibfield{author}{\bibinfo{person}{Shen Li}, \bibinfo{person}{Yanli Zhao}, \bibinfo{person}{Rohan Varma}, \bibinfo{person}{Omkar Salpekar}, \bibinfo{person}{Pieter Noordhuis}, \bibinfo{person}{Teng Li}, \bibinfo{person}{Adam Paszke}, \bibinfo{person}{Jeff Smith}, \bibinfo{person}{Brian Vaughan}, \bibinfo{person}{Pritam Damania}, {and} \bibinfo{person}{Soumith Chintala}.} \bibinfo{year}{2020}\natexlab{}.
\newblock \showarticletitle{PyTorch Distributed: Experiences on Accelerating Data Parallel Training}.
\newblock \bibinfo{journal}{\emph{Proc. {VLDB} Endow.}} \bibinfo{volume}{13}, \bibinfo{number}{12} (\bibinfo{year}{2020}), \bibinfo{pages}{3005--3018}.
\newblock


\bibitem[Lightning(2023)]%
        {WelcomePyTorchLightning}
\bibfield{author}{\bibinfo{person}{PyTorch Lightning}.} \bibinfo{year}{2023}\natexlab{}.
\newblock \bibinfo{title}{Welcome to {{PyTorch Lightning}} \textemdash{} {{PyTorch Lightning}} 2.1.0 Documentation}.
\newblock \bibinfo{howpublished}{https://lightning.ai/docs/pytorch/stable/}.
\newblock


\bibitem[Lin et~al\mbox{.}(2022)]%
        {lin2021m6}
\bibfield{author}{\bibinfo{person}{Junyang Lin}, \bibinfo{person}{An Yang}, \bibinfo{person}{Jinze Bai}, \bibinfo{person}{Chang Zhou}, \bibinfo{person}{Le Jiang}, \bibinfo{person}{Xianyan Jia}, \bibinfo{person}{Ang Wang}, \bibinfo{person}{Jie Zhang}, \bibinfo{person}{Yong Li}, \bibinfo{person}{Wei Lin}, \bibinfo{person}{Jingren Zhou}, {and} \bibinfo{person}{Hongxia Yang}.} \bibinfo{year}{2022}\natexlab{}.
\newblock \bibinfo{title}{M6-10T: A Sharing-Delinking Paradigm for Efficient Multi-Trillion Parameter Pretraining}.
\newblock
\newblock
\urldef\tempurl%
\url{https://openreview.net/forum?id=TXqemS7XEH}
\showURL{%
\tempurl}


\bibitem[Maurya et~al\mbox{.}(2021)]%
        {VELOC-MASCOTS21}
\bibfield{author}{\bibinfo{person}{Avinash Maurya}, \bibinfo{person}{Bogdan Nicolae}, \bibinfo{person}{Mustafa Rafique}, \bibinfo{person}{Thierry Tonellot}, {and} \bibinfo{person}{Franck Cappello}.} \bibinfo{year}{2021}\natexlab{}.
\newblock \showarticletitle{{Towards Efficient I/O Scheduling for Collaborative Multi-Level Checkpointing}}. In \bibinfo{booktitle}{\emph{{MASCOTS'21: The 29th IEEE International Symposium on the Modeling, Analysis, and Simulation of Computer and Telecommunication Systems}}}. \bibinfo{publisher}{IEEE}, \bibinfo{address}{Virtual, Portugal}, \bibinfo{pages}{1--8}.
\newblock
\urldef\tempurl%
\url{https://hal.inria.fr/hal-03344362}
\showURL{%
\tempurl}


\bibitem[Maurya et~al\mbox{.}(2022)]%
        {VELOCGPU-HiPC22}
\bibfield{author}{\bibinfo{person}{Avinash Maurya}, \bibinfo{person}{Bogdan Nicolae}, \bibinfo{person}{M.~Mustafa Rafique}, \bibinfo{person}{Amr~M. Elsayed}, \bibinfo{person}{Thierry Tonellot}, {and} \bibinfo{person}{Franck Cappello}.} \bibinfo{year}{2022}\natexlab{}.
\newblock \showarticletitle{{Towards Efficient Cache Allocation for High-Frequency Checkpointing}}. In \bibinfo{booktitle}{\emph{{HiPC'22: The 29th IEEE International Conference on High Performance Computing, Data, and Analytics}}}. \bibinfo{publisher}{IEEE}, \bibinfo{address}{Bangalore, India}, \bibinfo{pages}{262--271}.
\newblock


\bibitem[Maurya et~al\mbox{.}(2023b)]%
        {GPUPrefetch-HPDC23}
\bibfield{author}{\bibinfo{person}{Avinash Maurya}, \bibinfo{person}{Mustafa Rafique}, \bibinfo{person}{Thierry Tonellot}, \bibinfo{person}{Hussain AlSalem}, \bibinfo{person}{Franck Cappello}, {and} \bibinfo{person}{Bogdan Nicolae}.} \bibinfo{year}{2023}\natexlab{b}.
\newblock \showarticletitle{GPU-Enabled Asynchronous Multi-level Checkpoint Caching and Prefetching}. In \bibinfo{booktitle}{\emph{HPDC'23: The 32nd International Symposium on High-Performance Parallel and Distributed Computing}}. \bibinfo{publisher}{ACM}, \bibinfo{address}{Orlando, USA}, \bibinfo{pages}{73--85}.
\newblock
\urldef\tempurl%
\url{https://hal.inria.fr/hal-04119928}
\showURL{%
\tempurl}


\bibitem[Maurya et~al\mbox{.}(2023a)]%
        {LineageComp-HIPC23}
\bibfield{author}{\bibinfo{person}{Avinash Maurya}, \bibinfo{person}{M.~Mustafa Rafique}, \bibinfo{person}{Franck Cappello}, {and} \bibinfo{person}{Bogdan Nicolae}.} \bibinfo{year}{2023}\natexlab{a}.
\newblock \showarticletitle{Towards Efficient I/O Pipelines using Accumulated Compression}. In \bibinfo{booktitle}{\emph{HIPC’23: 30th IEEE International Conference on High Performance Computing, Data, and Analytics}}. \bibinfo{publisher}{IEEE}, \bibinfo{address}{Goa, India}, \bibinfo{pages}{256--265}.
\newblock


\bibitem[{Microsoft}(2023)]%
        {ziqiwangOptimizeCheckpointPerformance2023}
\bibfield{author}{\bibinfo{person}{{Microsoft}}.} \bibinfo{year}{2023}\natexlab{}.
\newblock \bibinfo{title}{Optimize {{Checkpoint Performance}} for {{Large Models}} - {{Azure Machine Learning}}}.
\newblock \bibinfo{howpublished}{\url{https://learn.microsoft.com/en-us/azure/machine-learning/reference-checkpoint-performance-for-large-models}}.
\newblock


\bibitem[Mohan et~al\mbox{.}(2021)]%
        {mohanCheckFreqFrequentFineGrained}
\bibfield{author}{\bibinfo{person}{Jayashree Mohan}, \bibinfo{person}{Amar Phanishayee}, {and} \bibinfo{person}{Vijay Chidambaram}.} \bibinfo{year}{2021}\natexlab{}.
\newblock \showarticletitle{{CheckFreq}: Frequent, {Fine-Grained} {DNN} Checkpointing}. In \bibinfo{booktitle}{\emph{FAST'21: The 19th USENIX Conference on File and Storage Technologies}}. \bibinfo{publisher}{USENIX Association}, \bibinfo{address}{Boston, USA}, \bibinfo{pages}{203--216}.
\newblock
\showISBNx{978-1-939133-20-5}


\bibitem[Nicolae et~al\mbox{.}(2020)]%
        {nicolaeDeepFreezeScalableAsynchronous2020}
\bibfield{author}{\bibinfo{person}{Bogdan Nicolae}, \bibinfo{person}{Jiali Li}, \bibinfo{person}{Justin~M. Wozniak}, \bibinfo{person}{George Bosilca}, \bibinfo{person}{Matthieu Dorier}, {and} \bibinfo{person}{Franck Cappello}.} \bibinfo{year}{2020}\natexlab{}.
\newblock \showarticletitle{{{DeepFreeze}}: {{Towards Scalable Asynchronous Checkpointing}} of {{Deep Learning Models}}}. In \bibinfo{booktitle}{\emph{CCGrid'20: The 20th {{International Symposium}} on {{Cluster}}, {{Cloud}} and {{Internet Computing}}}}. \bibinfo{publisher}{{IEEE/ACM}}, \bibinfo{address}{{Melbourne, Australia}}, \bibinfo{pages}{172--181}.
\newblock
\showISBNx{978-1-72816-095-5}


\bibitem[Nicolae et~al\mbox{.}(2019)]%
        {nicolaeVeloCHighPerformance2019}
\bibfield{author}{\bibinfo{person}{Bogdan Nicolae}, \bibinfo{person}{Adam Moody}, \bibinfo{person}{Elsa Gonsiorowski}, \bibinfo{person}{Kathryn Mohror}, {and} \bibinfo{person}{Franck Cappello}.} \bibinfo{year}{2019}\natexlab{}.
\newblock \showarticletitle{{{VeloC}}: {{Towards High Performance Adaptive Asynchronous Checkpointing}} at {{Large Scale}}}. In \bibinfo{booktitle}{\emph{IPDPS'19: {{IEEE International Parallel}} and {{Distributed Processing Symposium}}}}. \bibinfo{publisher}{IEEE}, \bibinfo{address}{Rio de Janeiro, Brazil}, \bibinfo{pages}{911--920}.
\newblock
\showISSN{1530-2075}


\bibitem[Nukada et~al\mbox{.}(2011)]%
        {nukada2011nvcr}
\bibfield{author}{\bibinfo{person}{Akira Nukada}, \bibinfo{person}{Hiroyuki Takizawa}, {and} \bibinfo{person}{Satoshi Matsuoka}.} \bibinfo{year}{2011}\natexlab{}.
\newblock \showarticletitle{NVCR: A transparent checkpoint-restart library for NVIDIA CUDA}. In \bibinfo{booktitle}{\emph{IPDPS'11: Proceedings of the International Symposium on Parallel and Distributed Processing Workshops and Phd Forum}}. \bibinfo{publisher}{IEEE}, \bibinfo{address}{Anchorage, AK, USA}, \bibinfo{pages}{104--113}.
\newblock


\bibitem[Parasyris et~al\mbox{.}(2020)]%
        {parasyris2020checkpoint-fti-gpu}
\bibfield{author}{\bibinfo{person}{Konstantinos Parasyris}, \bibinfo{person}{Kai Keller}, \bibinfo{person}{Leonardo Bautista-Gomez}, {and} \bibinfo{person}{Osman Unsal}.} \bibinfo{year}{2020}\natexlab{}.
\newblock \showarticletitle{Checkpoint restart support for heterogeneous hpc applications}. In \bibinfo{booktitle}{\emph{CCGRID'20: The International Symposium on Cluster, Cloud and Internet Computing (CCGRID)}}. \bibinfo{publisher}{IEEE/ACM}, \bibinfo{address}{Melbourne, Australia}, \bibinfo{pages}{242--251}.
\newblock


\bibitem[Potluri et~al\mbox{.}(2013)]%
        {potluri2013efficient}
\bibfield{author}{\bibinfo{person}{Sreeram Potluri}, \bibinfo{person}{Khaled Hamidouche}, \bibinfo{person}{Akshay Venkatesh}, \bibinfo{person}{Devendar Bureddy}, {and} \bibinfo{person}{Dhabaleswar~K Panda}.} \bibinfo{year}{2013}\natexlab{}.
\newblock \showarticletitle{Efficient inter-node MPI communication using GPUDirect RDMA for InfiniBand clusters with NVIDIA GPUs}. In \bibinfo{booktitle}{\emph{ICPP'13: The International Conference on Parallel Processing}}. \bibinfo{publisher}{IEEE}, \bibinfo{address}{Lyon, France}, \bibinfo{pages}{80--89}.
\newblock


\bibitem[PyTorch(2024)]%
        {TorchSnapshot}
\bibfield{author}{\bibinfo{person}{PyTorch}.} \bibinfo{year}{2024}\natexlab{}.
\newblock \bibinfo{title}{Welcome to the TorchSnapshot documentation}.
\newblock \bibinfo{howpublished}{\url{https://pytorch.org/torchsnapshot/stable/}}.
\newblock


\bibitem[PyTorch-Lightning(2024)]%
        {LightningAsyncCheckpointIO}
\bibfield{author}{\bibinfo{person}{PyTorch-Lightning}.} \bibinfo{year}{2024}\natexlab{}.
\newblock \bibinfo{title}{AsyncCheckpointIO-- PyTorch Lightning}.
\newblock
\newblock
\newblock
\shownote{\url{https://lightning.ai/docs/pytorch/stable/api/lightning.pytorch.plugins.io.AsyncCheckpointIO.html}}.


\bibitem[Rajbhandari et~al\mbox{.}(2020)]%
        {rajbhandariZeROMemoryOptimizations2020}
\bibfield{author}{\bibinfo{person}{Samyam Rajbhandari}, \bibinfo{person}{Jeff Rasley}, \bibinfo{person}{Olatunji Ruwase}, {and} \bibinfo{person}{Yuxiong He}.} \bibinfo{year}{2020}\natexlab{}.
\newblock \bibinfo{title}{{{ZeRO}}: {{Memory Optimizations Toward Training Trillion Parameter Models}}}.
\newblock
\newblock
\showeprint[arxiv]{1910.02054}~[cs, stat]


\bibitem[Rajbhandari et~al\mbox{.}(2021)]%
        {rajbhandari2021zero}
\bibfield{author}{\bibinfo{person}{Samyam Rajbhandari}, \bibinfo{person}{Olatunji Ruwase}, \bibinfo{person}{Jeff Rasley}, \bibinfo{person}{Shaden Smith}, {and} \bibinfo{person}{Yuxiong He}.} \bibinfo{year}{2021}\natexlab{}.
\newblock \showarticletitle{ZeRO-infinity: breaking the GPU memory wall for extreme scale deep learning}. In \bibinfo{booktitle}{\emph{SC'21: The International Conference for High Performance Computing, Networking, Storage and Analysis}}. \bibinfo{publisher}{ACM}, \bibinfo{address}{St. Louis, Missouri}, Article \bibinfo{articleno}{59}, \bibinfo{numpages}{14}~pages.
\newblock
\showISBNx{9781450384421}


\bibitem[Rasley et~al\mbox{.}(2020)]%
        {rasleyDeepSpeedSystemOptimizations2020}
\bibfield{author}{\bibinfo{person}{Jeff Rasley}, \bibinfo{person}{Samyam Rajbhandari}, \bibinfo{person}{Olatunji Ruwase}, {and} \bibinfo{person}{Yuxiong He}.} \bibinfo{year}{2020}\natexlab{}.
\newblock \showarticletitle{{{DeepSpeed}}: {{System Optimizations Enable Training Deep Learning Models}} with {{Over}} 100 {{Billion Parameters}}}. In \bibinfo{booktitle}{\emph{KDD'20: The 26th {{SIGKDD International Conference}} on {{Knowledge Discovery}} \& {{Data Mining}}}}. \bibinfo{publisher}{{ACM}}, \bibinfo{address}{{Virtual Event CA USA}}, \bibinfo{pages}{3505--3506}.
\newblock
\showISBNx{978-1-4503-7998-4}


\bibitem[Ruder(2017)]%
        {ruder2016overview}
\bibfield{author}{\bibinfo{person}{Sebastian Ruder}.} \bibinfo{year}{2017}\natexlab{}.
\newblock \bibinfo{title}{An overview of gradient descent optimization algorithms}.
\newblock
\newblock
\showeprint[arxiv]{1609.04747}~[cs.LG]


\bibitem[Schwan et~al\mbox{.}(2003)]%
        {schwan2003lustre}
\bibfield{author}{\bibinfo{person}{Philip Schwan} {et~al\mbox{.}}} \bibinfo{year}{2003}\natexlab{}.
\newblock \showarticletitle{Lustre: Building a file system for 1000-node clusters}. In \bibinfo{booktitle}{\emph{Proceedings of the 2003 Linux symposium}}, Vol.~\bibinfo{volume}{2003}. \bibinfo{publisher}{Linux symposium}, \bibinfo{address}{Ontario, Canada}, \bibinfo{pages}{380--386}.
\newblock


\bibitem[Shoeybi et~al\mbox{.}(2020)]%
        {shoeybiMegatronLMTrainingMultiBillion2020}
\bibfield{author}{\bibinfo{person}{Mohammad Shoeybi}, \bibinfo{person}{Mostofa Patwary}, \bibinfo{person}{Raul Puri}, \bibinfo{person}{Patrick LeGresley}, \bibinfo{person}{Jared Casper}, {and} \bibinfo{person}{Bryan Catanzaro}.} \bibinfo{year}{2020}\natexlab{}.
\newblock \bibinfo{title}{Megatron-{{LM}}: {{Training Multi-Billion Parameter Language Models Using Model Parallelism}}}.
\newblock
\newblock
\showeprint[arxiv]{1909.08053}~[cs]


\bibitem[Simonyan and Zisserman(2015)]%
        {vgg-paper}
\bibfield{author}{\bibinfo{person}{Karen Simonyan} {and} \bibinfo{person}{Andrew Zisserman}.} \bibinfo{year}{2015}\natexlab{}.
\newblock \bibinfo{title}{Very Deep Convolutional Networks for Large-Scale Image Recognition}.
\newblock
\newblock
\showeprint[arxiv]{1409.1556}~[cs.CV]


\bibitem[Song et~al\mbox{.}(2023)]%
        {songDeepSpeed4ScienceInitiativeEnabling2023}
\bibfield{author}{\bibinfo{person}{Shuaiwen~Leon Song}, \bibinfo{person}{Bonnie Kruft}, \bibinfo{person}{Minjia Zhang}, \bibinfo{person}{Conglong Li}, \bibinfo{person}{Shiyang Chen}, {et~al\mbox{.}}} \bibinfo{year}{2023}\natexlab{}.
\newblock \bibinfo{title}{{{DeepSpeed4Science Initiative}}: {{Enabling Large-Scale Scientific Discovery}} through {{Sophisticated AI System Technologies}}}.
\newblock
\newblock
\showeprint[arxiv]{2310.04610}~[cs]


\bibitem[Takizawa et~al\mbox{.}(2009)]%
        {takizawa2009checuda}
\bibfield{author}{\bibinfo{person}{Hiroyuki Takizawa}, \bibinfo{person}{Katsuto Sato}, \bibinfo{person}{Kazuhiko Komatsu}, {and} \bibinfo{person}{Hiroaki Kobayashi}.} \bibinfo{year}{2009}\natexlab{}.
\newblock \showarticletitle{CheCUDA: A Checkpoint/Restart Tool for CUDA Applications}. In \bibinfo{booktitle}{\emph{PDCAT'09: The International Conference on Parallel and Distributed Computing, Applications and Technologies}}. \bibinfo{publisher}{IEEE}, \bibinfo{address}{Higashi-Hiroshima, Japan}, \bibinfo{pages}{408--413}.
\newblock


\bibitem[Touvron et~al\mbox{.}(2023)]%
        {touvronLlamaOpenFoundation2023}
\bibfield{author}{\bibinfo{person}{Hugo Touvron}, \bibinfo{person}{Louis Martin}, \bibinfo{person}{Kevin Stone}, \bibinfo{person}{Peter Albert}, \bibinfo{person}{Amjad Almahairi}, {et~al\mbox{.}}} \bibinfo{year}{2023}\natexlab{}.
\newblock \bibinfo{title}{Llama 2: {{Open Foundation}} and {{Fine-Tuned Chat Models}}}.
\newblock
\newblock
\showeprint[arxiv]{2307.09288}~[cs]


\bibitem[Wang et~al\mbox{.}(2023b)]%
        {wang2023reliableREFT}
\bibfield{author}{\bibinfo{person}{Yuxin Wang}, \bibinfo{person}{Shaohuai Shi}, \bibinfo{person}{Xin He}, \bibinfo{person}{Zhenheng Tang}, \bibinfo{person}{Xinglin Pan}, \bibinfo{person}{Yang Zheng}, \bibinfo{person}{Xiaoyu Wu}, \bibinfo{person}{Amelie~Chi Zhou}, \bibinfo{person}{Bingsheng He}, {and} \bibinfo{person}{Xiaowen Chu}.} \bibinfo{year}{2023}\natexlab{b}.
\newblock \bibinfo{title}{Reliable and Efficient In-Memory Fault Tolerance of Large Language Model Pretraining}.
\newblock
\newblock
\showeprint[arxiv]{2310.12670}~[cs.DC]


\bibitem[Wang et~al\mbox{.}(2023a)]%
        {wang2023gemini}
\bibfield{author}{\bibinfo{person}{Zhuang Wang}, \bibinfo{person}{Zhen Jia}, \bibinfo{person}{Shuai Zheng}, \bibinfo{person}{Zhen Zhang}, \bibinfo{person}{Xinwei Fu}, \bibinfo{person}{T.~S.~Eugene Ng}, {and} \bibinfo{person}{Yida Wang}.} \bibinfo{year}{2023}\natexlab{a}.
\newblock \showarticletitle{GEMINI: Fast Failure Recovery in Distributed Training with In-Memory Checkpoints}. In \bibinfo{booktitle}{\emph{SOSP'23: The Proceedings of the 29th Symposium on Operating Systems Principles}} \emph{(\bibinfo{series}{SOSP '23})}. \bibinfo{publisher}{ACM}, \bibinfo{address}{Koblenz, Germany}, \bibinfo{pages}{364–381}.
\newblock
\showISBNx{9798400702297}


\bibitem[Workshop et~al\mbox{.}(2023)]%
        {workshopBLOOM176BParameterOpenAccess2023}
\bibfield{author}{\bibinfo{person}{BigScience Workshop}, \bibinfo{person}{Teven~Le Scao}, \bibinfo{person}{Angela Fan}, \bibinfo{person}{Christopher Akiki}, \bibinfo{person}{Ellie Pavlick}, \bibinfo{person}{Suzana Ili{\'c}}, {et~al\mbox{.}}} \bibinfo{year}{2023}\natexlab{}.
\newblock \bibinfo{title}{{{BLOOM}}: {{A 176B-Parameter Open-Access Multilingual Language Model}}}.
\newblock
\newblock
\showeprint[arxiv]{2211.05100}~[cs]


\bibitem[Wu et~al\mbox{.}(2023)]%
        {wu2023transom}
\bibfield{author}{\bibinfo{person}{Baodong Wu}, \bibinfo{person}{Lei Xia}, \bibinfo{person}{Qingping Li}, \bibinfo{person}{Kangyu Li}, \bibinfo{person}{Xu Chen}, \bibinfo{person}{Yongqiang Guo}, \bibinfo{person}{Tieyao Xiang}, \bibinfo{person}{Yuheng Chen}, {and} \bibinfo{person}{Shigang Li}.} \bibinfo{year}{2023}\natexlab{}.
\newblock \bibinfo{title}{TRANSOM: An Efficient Fault-Tolerant System for Training LLMs}.
\newblock
\newblock
\showeprint[arxiv]{2310.10046}~[cs.DC]


\bibitem[Zeng et~al\mbox{.}(2023)]%
        {zeng2022glm}
\bibfield{author}{\bibinfo{person}{Aohan Zeng}, \bibinfo{person}{Xiao Liu}, \bibinfo{person}{Zhengxiao Du}, \bibinfo{person}{Zihan Wang}, \bibinfo{person}{Hanyu Lai}, \bibinfo{person}{Ming Ding}, \bibinfo{person}{Zhuoyi Yang}, \bibinfo{person}{Yifan Xu}, \bibinfo{person}{Wendi Zheng}, \bibinfo{person}{Xiao Xia}, \bibinfo{person}{Weng~Lam Tam}, \bibinfo{person}{Zixuan Ma}, \bibinfo{person}{Yufei Xue}, \bibinfo{person}{Jidong Zhai}, \bibinfo{person}{Wenguang Chen}, \bibinfo{person}{Peng Zhang}, \bibinfo{person}{Yuxiao Dong}, {and} \bibinfo{person}{Jie Tang}.} \bibinfo{year}{2023}\natexlab{}.
\newblock \bibinfo{title}{GLM-130B: An Open Bilingual Pre-trained Model}.
\newblock
\newblock
\showeprint[arxiv]{2210.02414}~[cs.CL]


\end{thebibliography}

\end{document}